\documentclass[iop]{emulateapj}
\slugcomment{{\it Recieved 2010 April 20; Accepted 2011 May 9}}
\usepackage{natbib}
\begin{document}
\title{X-ray and UV emission from the recurrent nova RS Ophiuchi in quiescence: Signatures of accretion and shocked gas}
\author{T. Nelson, K. Mukai\altaffilmark{1,2}, M. Orio\altaffilmark{3,4}, G. J. M. Luna\altaffilmark{5}, J. L. Sokoloski\altaffilmark{6}}
\altaffiltext{1}{CRESST and X-ray Astrophysics Laboratory NASA/GSFC, Greenbelt, MD 20771, USA}
\altaffiltext{2}{Center for Space Science and Technology, University of Maryland Baltimore County, 1000 Hilltop Circle, Baltimore MD 21250, USA}
\altaffiltext{3}{INAF - Osservatorio Astronomico di Padova, vicolo Osservatorio, 5, I-35122 
Padova, Italy}
\altaffiltext{4}{Department of Astronomy, University of Wisconsin-Madison, 475 N Charter St, 
Madison, WI, 53706, USA}
\altaffiltext{5}{Harvard-Smithsonian Center for Astrophysics, 60 Garden St. MS15, Cambridge, MS 02138, USA}
\altaffiltext{6}{Columbia Astrophysics Laboratory, 550 W. 220th St, 1027 Pupin Hall, Columbia University, New York, NY 10027, USA}
\email{thomas.nelson@nasa.gov}

\begin{abstract}
RS Ophiuchi is a recurrent nova system that experiences outbursts every $\sim$20 years, implying accretion at a high rate onto a massive white dwarf.  However, previous X-ray observations of the system in quiescence have detected only faint emission that is difficult to reconcile with the high accretion rate predicted by nova theory for such frequent outbursts. Here, we use new \textit{Chandra} and \textit{XMM-Newton} observations obtained 537 and 744 days after the 2006 outburst to constrain both the accretion rate onto the white dwarf and the properties of the nova ejecta at these times.  We detect low level UV variability with the \textit{XMM-Newton} Optical Monitor on day 744 that is consistent with accretion disk flickering, and use this to place a lower limit on the accretion rate.  The X-ray spectra in both observations are well described by a two component thermal plasma model.  The first component originates in the nova shell, which can emit X-rays for up to a decade after the outburst.  The other component likely arises in the accretion disk boundary layer, and can be equally well fit by a single temperature plasma or a cooling flow model.  Although the flux of the single temperature model implies an accretion rate that is 40 times lower than theoretical predictions for RS Oph, the best fit cooling flow model implies $\dot{M}$ $<$ 1.2 $\times$ 10$^{-8}$  M$_{\odot}$ yr$^{-1}$ 537 days after the outburst, which is within a factor of 2 of the theoretical accretion rate required to power an outburst every 20 years.  Furthermore, we place an upper limit on the accretion rate through an optically thick region of the boundary layer of 2.0 $\times$ 10$^{-8}$  M$_{\odot}$ yr$^{-1}$. Thus, these new quiescence data are consistent with the accretion rate expectations of nova theory.  Finally, we discuss the possible origins of the low temperature associated with the accretion component, which is a factor of 10 lower than in T CrB, an otherwise similar recurrent nova.
\end{abstract}

\section{Introduction}
\label{intro}
Recurrent novae are a rare subclass of cataclysmic variable stars (CVs; white dwarfs accreting material from a binary companion) in which more than one classical nova-type outburst has been observed.  These outbursts are believed to be due to a thermonuclear runaway on the surface of the white dwarf, which commences once a critical pressure is reached at the base of the shell of accreted material.  All classical novae are expected to recur on timescales from 100,000 years to just a few decades.  The most important physical parameters controlling this recurrence timescale are the white dwarf mass, and the mass accretion rate from the secondary \citep[see e.g.][]{Yaron05}. RS Oph is a particularly prolific recurrent nova, with detected outbursts in 1898, 1907, 1933, 1945, 1958, 1967, 1985 and 2006 \citep{Schaefer09}.  The short time between outbursts ($\sim$20 yrs) suggests that RS Oph hosts a massive white dwarf accreting material at a high rate. 

Observations of RS Oph during its 2006 outburst allowed direct observational constraints to be placed on the mass of the white dwarf. \citet{Nelson08a} presented high resolution X-ray spectra obtained during the supersoft phase (the phase where the still burning hydrogen shell is directly observable), and found a very high effective temperature for the source ($\sim$800,000 K), consistent with shell burning on a white dwarf of at least 1.2 M$_{\odot}$ \citep[see e.g. ][]{Paczynski71,Nomoto07}.  Subsequent analysis of the {\it Swift} data obtained during the same epoch by \citet{Osborne11} led those authors to also conclude that the white dwarf mass must be more than 1.2 M$_{\odot}$.  Furthermore, \citet{Sokoloski06} estimated the mass of the ejected envelope based on the hard X-ray light curve of the outburst, and found that no more than a few 10$^{-7}$ M$_{\odot}$ could have been ejected.  This is consistent with the predicted ejecta mass for a nova event on a 1.4 M$_{\odot}$ WD \citep[$\sim$2 $\times$ 10$^{-7}$ M$_{\odot}$, ][]{Yaron05}. Thus, the white dwarf in RS Oph is extremely massive, in agreement with the first expectation for systems with frequent nova outbursts. 

However, there is less evidence of the high accretion rate required to power an outburst every $\sim$20 yrs.  Cataclysmic variables are well known X-ray sources, and are understood to be powered by accretion.  For a slowly rotating white dwarf, approximately half of the available accretion luminosity is emitted in the accretion disk boundary layer at X-ray wavelengths \citep[see e.g.][]{Shakura73,Lynden-Bell74}.  Therefore, a measurement of the X-ray luminosity can be used to estimate the accretion rate as follows:
\begin{equation}
\dot{M} = \frac{2L_{X}R_{WD}}{GM_{WD}} 
\end{equation} where $\dot{M}$ is the mass accretion rate, L$_{X}$ is the X-ray luminosity, and R$_{WD}$ and M$_{WD}$ and the radius and mass of the white dwarf, respectively.  The same method can be applied to existing X-ray observations of RS Oph.  \citet{Orio93} and \citet{Orio01} presented \textit{ROSAT} PSPC observations of RS Oph carried out 6 and 7 years after the 1985 outburst, respectively.  In those observations, the system was detected as rather faint, soft X-ray source (all photons with energies $<$1.5 keV), which varied in count rate between the two observations by a factor of 3.  The derived luminosities in the \textit{ROSAT} energy range (0.2--2.4 keV), based on best fitting thermal plasma models and assuming a distance to the system of 1.6 kpc, were 3 $\times$ 10$^{31}$ to 5 $\times$ 10$^{32}$ erg s$^{-1}$.  Using Equation 1, and assuming a white dwarf mass of 1.3 M$_{\odot}$ and radius 3 $\times$ 10$^{8}$ cm \citep{Althaus05}, this low luminosity implies a mass accretion rate in the range 10$^{-12}$ to a few 10$^{-11}$ M$_{\odot}$ yr$^{-1}$.  A similar range of luminosities, and hence accretion rates, was found by \citet{Mukai09} using a 35 ks \textit{ASCA} observation of the system carried out in 1997 ($\sim$12 yrs after the 1985 outburst).  Such low rates are completely at odds with the short nova outburst recurrence time ($\sim$20 years), which requires $\dot{M}$ $>$10$^{-8}$ M$_{\odot}$ yr$^{-1}$ for a 1.4 M$_{\odot}$ WD, and $>$10$^{-7}$ M$_{\odot}$ yr$^{-1}$ if the mass of the WD is only 1.25 M$_{\odot}$ \citep{Yaron05}.  

RS Oph is also categorized as a symbiotic star i.e. a star whose binary nature is revealed through the presence of two distinct components in its spectrum, one hot and one cool.  Most of these systems are comprised of a white dwarf accreting material from a red giant companion in a wide binary, with orbital periods in the range of a few years to a few decades. Symbiotic stars can therefore be considered larger scale cousins of the cataclysmic variables.  The mode of accretion onto the white dwarf has not been definitively identified in symbiotic stars.  A small number of systems, including RS Oph, exhibit strong optical flickering (stochastic brightness fluctuations on timescales of minutes to hours), which is also seen in CVs with disks \citep{Sokoloski01}.   These brightness fluctuations are thought to be due to the turbulent motion of material as it passes through the disk \citep{Bruch92}, and their presence is taken as evidence of an accretion disk in those systems.  \citet{Sokoloski03} reported the presence of minute time scale variability in the He II $\lambda$4686 line in RS Oph in quiescence.  The high ionization potential of this ion (56 eV) requires a source of soft X-rays, most likely the accretion disk boundary layer.  Additionally, the launching of radio jets during the 2006 outburst \citep{Sokoloski08,Rupen08} can also be considered evidence of the presence of an accretion disk, since almost all mechanisms for producing jets require a disk to anchor the magnetic fields that collimate the outflow.  We refer the reader to \citet{Wynn09} for a detailed discussion of the possible modes of accretion in RS Oph.

Although the presence of a disk in RS Oph seems likely, the mechanism by which material is fed into it is unknown.  Most of the giants in symbiotic stars do not appear to fill their Roche Lobes, although a number of systems show near-infrared ellipsoidal variations, suggesting that they are at least partially tidally distorted \citep{Mikolajewska07}.   If Roche lobe overflow (RLO) is not occurring, it is normally assumed that material is fed to the white dwarf through Bondi-Hoyle accretion of the red giant wind.  Since this mode of accretion is much less efficient than RLO, it is unclear that it can provide enough mass to produce an outburst in RS Oph every 20 years.  

No ellipsoidal variations have been observed in RS Oph, which is generally assumed to indicate that it does not fill its Roche lobe. If this is the case, then the accretion disk must be fed by some mechanism other than RLO.  Work by \citet{Podsiadlowski07} explores one such alternative.  Those authors presented a scenario dubbed ``wind Roche lobe overflow", where the stellar wind of the giant fills the Roche lobe, even though its photosphere (determined from mass-luminosity-radius relationships) does not.  The wind is gravitationally focused towards the L1 point, where it is then captured by the potential well of the white dwarf.  At this point, the material can form an accretion disk, as in RLO systems.  This mode of mass transfer can be up to 100 times more efficient than Bondi-Hoyle accretion.  If such mass transfer is occurring in RS Oph, it can feed a disk, and the white dwarf, at a sufficiently high rate to power the observed outburst frequency without the red giant filling its Roche lobe.

In this paper, we present new deep, quiescent-state X-ray observations of RS Oph, carried out with \textit{Chandra} and \textit{XMM-Newton} 537 and 744 days after the 2006 outburst, respectively.  These observations shed light on the origins of the X-ray emission in the system, and reveal new details about the accretion of material onto the white dwarf component.  In Section~\ref{data} we describe the observations and the data reduction techniques used to obtain lightcurves and spectra for analysis.  We present our timing analysis in Section~\ref{lightcurves} and our spectral analysis and best fitting models in Section~\ref{spectra}.  We discuss the derived quiescence properties of RS Oph in the context of earlier X-ray observations of the system in Section~\ref{comparison}.  In Section~\ref{discussion}, we discuss our findings and their implications for understanding the physics of accretion in the RS Oph system.  Finally, a summary of our work is given in Section~\ref{conclusions}.

\begin{deluxetable*}{cccccc}[h!]
\tabletypesize{\footnotesize}
\tablecaption{New observations of RS Oph}
\tablewidth{0pt}
\tablehead{
\colhead{Date} & \colhead{Observatory} & \colhead{Instrument} & \colhead{ Net t$_{exp}$ (s)} & \colhead{Days from outburst}\tablenotemark{a} & \colhead{Flux (erg s$^{-1}$ cm$^{-2}$)}\tablenotemark{b}
}
\startdata
2007-08-04 06:11:26 & \textit{Chandra} & ACIS-S & 90993 & 537 & 5.2 $\times$ 10$^{-13}$\\
2008-02-26 16:57:59 & \textit{XMM-Newton} & EPIC-pn & 20780 & 744 & 2.1 $\times$ 10$^{-13}$\\
                    &                     & EPIC-MOS1 & 28168 &     & \\
		    &                     & EPIC-MOS2 & 28461 &     & 
\enddata
\tablenotetext{a}{We set $t_{0}$ = 2,453,779.33 after Narumi et al. (2006).}
\tablenotetext{b}{Observed flux, not corrected for absorption. Fluxes are in the energy range 0.3--10.0 keV}
\end{deluxetable*}

\section{Observations and Data Reduction}
\label{data}

The first of our observations of RS Oph was carried out on 2007 August 4 (537 days after the 2006 outburst began) with the ACIS-S camera onboard the \textit{Chandra} X-ray observatory, for a total exposure time of 91 ks (see Table 1).  This exposure was triggered as part of a target of opportunity (ToO) program to image jet activity in symbiotic stars.  Faint extended structure is in fact present in the data, and is presented in \citet{Luna09}.  The exposure was performed in FAINT mode, with the source positioned on the ACIS-S3 chip (one of two back-illuminated CCDs) which has the highest sensitivity to soft photons.  The data as downloaded from the archive had been reduced with the most recent calibration files available, so no further reprocessing of the data was necessary.  We extracted timing and spectral products using the \textit{Chandra} Interactive Analysis of Observations (CIAO) software, version 4.1.1.

Source lightcurves and spectra were extracted from a 6" circle centered on the source using the CIAO task {\tt dmextract}.  Background data were extracted from an annulus also centered on the source, with inner and outer radii of 10" and 20" respectively.  There was a total of 6468 good events in the source region, with 264 background counts, or 4\%  of the observed counts.  Spectral response files were generated using the CIAO tasks {\tt mkacisrmf} (response matrix) and {\tt mkarf} (effective area).  In order to facilitate the use of $\chi^{2}$ statistics, we grouped the spectra to give a minimum of thirty counts per bin.  Using PIMMS, we verified that our instrumental setup resulted in a pile up rate of less than 2\%.

We observed RS Oph for a second time on 2008 February 26 (day 744 after outburst), this time with the \textit{XMM-Newton} observatory.  All instruments onboard were turned on during the observation.  The European Photon Imaging Cameras (EPIC) MOS and pn instruments were all operated in full window mode, which exposes the entire field of view of the camera.  We reduced these data using the Science Analysis Software (SAS) v. 11.0.  First, we reran the pipeline routines in SAS in order to apply the most up to date calibration to the data.  The observation, which had a total duration of 35 ks, was affected by two periods of high background - one at the start of the observation, and a shorter interval near the end.  We screened these periods of high background from the event file by defining a new good time interval (GTI) file, which excludes time periods where the EPIC-pn (EPIC-MOS) 10--12 keV count rate is $>$0.4 (0.35) cts s$^{-1}$.  This reduced the total exposure time for each instrument (see Table 1), and also resulted in a low number of counts above 2 keV.  In order to improve the statistics for spectral fitting at high energies, we filtered the good time interval data by grade, retaining both single and double events i.e. PATTERN$<$=4, in the EPIC-pn data\footnote{In general, it is recommended that only events with grade 0 (i.e. singles only) are retained for spectral analysis of EPIC-pn data.  Using both singles and doubles can in principle increase the pile up rate, and reduce the energy resolution of the resulting spectrum. However RS Oph does not have a high enough count rate for this effect to be a concern.}.  For the MOS data, we retained events with grades $<=$ 12.  We masked bad pixels and CCD edge defects using the most conservative screening criterion, FLAG==0, as recommended for spectral analysis.  

Source lightcurves and spectra were extracted from a 30" circular region for all three EPIC instruments.  Background events in the two MOS datasets were extracted from an annular region of inner radius 35" and outer radius 100".  For the pn data, the background events were extracted from two rectangular regions at the same chip y coordinate, chosen to maximize the background area while extracting events with similar distance from the read out register, as recommended in the XMM-Newton EPIC calibration documentation. A total of 2047 good counts were extracted in the source region in the EPIC-pn data, with 74 of these (4\%) due to the background. The number of counts extracted from the MOS data was much lower - 630 (636) from MOS1 (MOS2), with 119 (109) background counts.  Spectral response files for all the EPIC spectra were created using the SAS tasks {\tt rmfgen} and {\tt arfgen}.  Again, we binned the spectra with a minimum of thirty counts per bin.  With PIMMS, we verified that the pile up rate in this case was less than 1\%.

The two reflection grating spectrograph (RGS) instruments were operated in standard spectroscopy mode, but no usable data was obtained due to the low count rate and we do not discuss them further.  The optical monitor (OM), operating in imaging mode, obtained images of the source in the V, U and UVM2 filters, with most exposures lasting 800s (the default for the standard mosaic pattern).  The complete dataset is comprised of 5 images in each of the V and U filters, and 11 in the UVM2 filter.  Note that the OM can also be operated in fast mode, in which the detector is continuously read out every 0.5s.  Use of this mode would be extremely useful in any future observations of RS Oph, since it provides much higher time resolution than standard imaging mode.  We analyzed the pipeline reduced images, which include photometry of the brightest sources in each image.  The pipeline carries out the following photometric procedure: source counts are extracted from a 16.5 arcsec radius circle, and background counts from a 18.3--20.7 arcsec annulus.  These are converted to count rates by dividing by the exposure time.  Next, these count rates are corrected for bad pixels and photon coincidence losses, which are well understood for the OM.  Finally, background counts are subtracted from the source annulus, based on the corrected count rates.  Magnitudes are then calculated using this corrected count rate and the zero-point defined for each filter in the XMM documentation.\footnote{http://xmm.esac.esa.int/sas/current/howtousesas.shtml}

\section{Timing Analysis}
\label{lightcurves}
The \textit{Chandra} lightcurves are presented in Figure 1, binned into 4000 s segments.  The most striking feature of these lightcurves is the rise in count rate during the first half of the observation, which increases from 0.05 to 0.08 c s$^{-1}$ in the 0.3--10 keV energy range.  We extracted X-ray lightcurves from each observation across the entire energy response of the instrument, and in three smaller energy bands: 0.3--1, 1--2 and 2--10 keV. Looking at the lightcurve behavior in the smaller energy bands, it is clear that the count rate increase occurs primarily at intermediate energies, between 1 and 2 keV (see Fig. 1, green data points).  The \textit{XMM-Newton} lightcurve is shown in Figure 2.  There are relatively more counts in the soft band compared to the medium and hard bands than in the \textit{Chandra} observation, although some of this effect can be explained by the different energy response of the EPIC-pn and ACIS-S instruments. However, that the 0.3--10 keV count rates are similar shows that the system must be intrinsically fainter during the \textit{XMM-Newton} observation, since \textit{XMM-Newton} has a much larger collecting area than \textit{Chandra}. We do not observe any significant count rate changes over the course of the \textit{XMM-Newton} observation, although we note that exposure time is shorter than the timescales over which such changes occurred in the \textit{Chandra} observation.

\begin{figure}
\begin{center}
\includegraphics[width=3.25in]{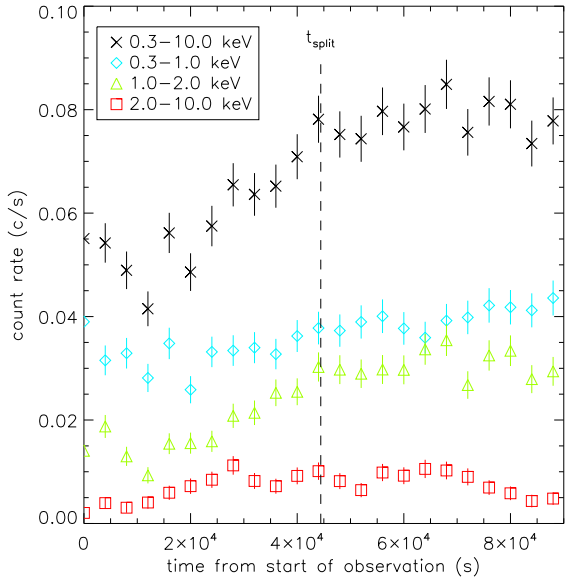}
\caption{ACIS-S lightcurves in various energy ranges, binned in 4000s intervals.  The vertical dashed line indicates the time used to split the observation into two spectra (see text).}
\end{center}
\end{figure}

\begin{figure}
\begin{center}
\includegraphics[width=3.25in]{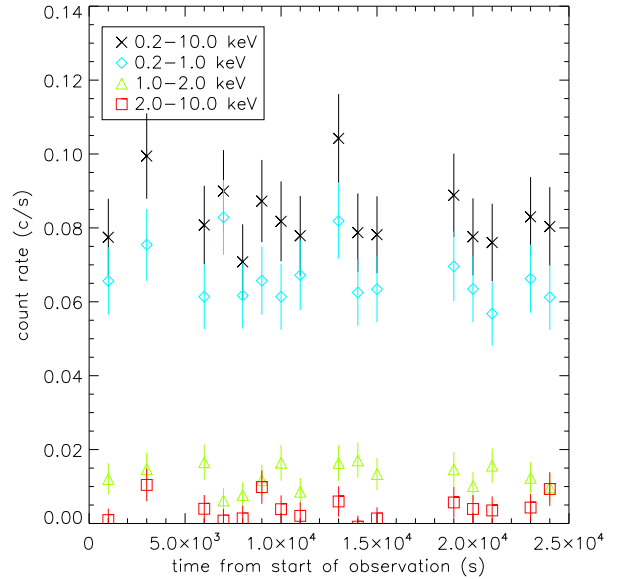}
\caption{XMM-Newton EPIC-pn lightcurve, binned in 1000s intervals.}
\end{center}
\end{figure}

The OM lightcurves in the U, V and UVM2 filters are shown in Figure 3, with no correction for interstellar reddening.  The $U-V$ color is consistent with the values reported by \citet{Henden08}. The UVM2 data show low-level variability from exposure to exposure. Following \citet{Sokoloski01}, we compared the standard deviation of the pipeline measured magnitudes with that expected if the observed changes in magnitude were due only to the errors (Poisson plus systematic) in the individual measurements.  This is the same as testing the hypothesis that the observed lightcurve is consistent with a source of constant flux using the $\chi^{2}$ statistic.  The expected standard deviation, $s_{exp}$, is found from the average of the observed errors on each measurement.  These are calculated by the XMM SAS pipeline, based on the Poisson errors of the observed source and background counts.  Taking the average value is appropriate here since the exposure time of each observation was the same. 

\begin{figure}[h!]
\begin{center}
\includegraphics[width=3.25in]{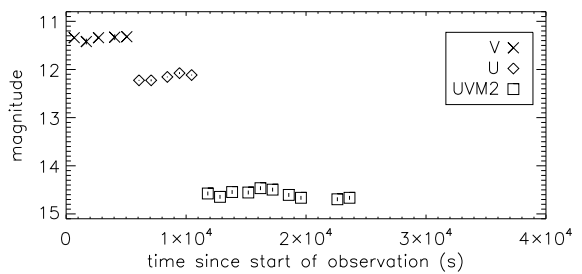}
\caption{Optical Monitor V, U and UVM2 lightcurves.  Magnitudes are in the standard Vega system. Note that the uncertainties on each point are smaller than the size of the symbols.}
\end{center}
\end{figure}

We find that the observed standard deviation of the UVM2 brightness (0.077 mags) is $\sim$2 times larger than that expected from the measured errors alone (0.037 mags),  a firm detection of variability.  This could be due to accretion induced flickering, which tends to be enhanced in bluer filters where the disk light dominates over the emission from the cooler companion.  If we work under the assumption that the observed fluctuations \textit{are} due to flickering, and hence accretion, we can use the amplitude of the flickering to place constraints on the accretion rate two years after the outburst.  We follow the method of \citet{Bruch92}, who assumed that the flickering source was 100\% modulated and that all other sources of light were constant.  If the amplitude of the flickering is $\Delta$$m$ and the mean observed magnitude over the course of the observations is $\bar{m}$, then the magnitude of the constant source, $m_{c}$ is given by
\begin{equation}
m_{c} = \bar{m} +0.5\Delta m 
\end{equation}
Assuming that the faintest data point in the lightcurve is due only to the flux of the constant source, $F_{c}$, and that the brightest data point is due to the constant source plus the flickering source ($F_{c} + F_{f}$), then it is straightforward to show that
\begin{equation}
\frac{F_{f}}{F_{c}} = 10^{0.4\Delta m} - 1.
\end{equation} 
We obtain both $\Delta m$ (0.23 mags) and $\bar{m}$ (14.56 mags) from the lightcurve.  The resulting ratio of $F_{f}$/$F_{c}$ is 0.24.  

In order to estimate the actual flux of the flickering source, the observed magnitude must first be corrected for the effects of interstellar extinction.  A number of different E(B-V) estimates are given in the literature, ranging from 0.4 to 0.73.  This spread in the estimate of E(B-V) is the dominant source of uncertainty in calculating the luminosity of the flickering source, since this corresponds to a difference in the magnitude correction of 2.63, or a change in the measured flux of a factor of $\sim$10.  Deciding which value to use depends somewhat on where the UV light is believed to originate.  Some intrinsic absorption of the accretion disk is likely, given that the white dwarf is embedded within the wind of the red giant.  Therefore, we calculated the flickering source flux for a range of reddening values.  First, we obtained the extinction coefficient in the UVM2 filter, $A_{2310}$, using the reddening law of \citet{Cardelli89}, for E(B-V) values of 0.4, 0.5, 0.6 and 0.7.  These values are presented in Table 2, along with $\bar{m}_{0}$, the reddening corrected mean magnitude for each value of E(B-V).  While the spread in E(B-V) is the largest source of uncertainty, an important source of error for a given E(B-V) value is the potential deviation from the assumed Cardelli et al. extinction law.  From the data presented in \citet{Fitzpatrick99}, we estimate that this effect contributes a 0.5 mag error into the extinction, and thus into the corrected mean magnitude.  This corresponds to an uncertainty in the derived flux of $\sim$68\%.  Next, we converted the $\bar{m}_{0}$ values into count rates, and then to fluxes, using the conversion factors outlined in the Optical Monitor calibration documentation, and the effective width of the UVM2 filter, 400 \AA.  Finally, we obtained the flux of the flickering component by multiplying by the ratio found using Equation 3.  These fluxes are presented in Table 2, and are in the range 1.5 $\times$ 10$^{-11}$ to 1.7 $\times$ 10$^{-10}$ erg s$^{-1}$ cm$^{-2}$, depending on the assumed value of E(B-V).

\begin{deluxetable}{ccccc}
\tabletypesize{\footnotesize}
\tablecaption{Accretion rate lower limits from UV flickering}
\tablewidth{0pt}
\tablehead{
\colhead{E(B-V)} & \colhead{A$_{2310}$} & \colhead{$\bar{m}_{0}$} & \colhead{$F_{f}$ (erg s$^{-1}$ cm$^{-2}$)} & \colhead{$\dot{M}$ (M$_{\odot, UV}$ yr$^{-1}$)}\tablenotemark{a}
}
\startdata
0.4 & 3.51 & 11.05 & 1.5 $\times$ 10$^{-11}$ & 2.3 $\times$ 10$^{-10}$ \\
0.5 & 4.38 & 10.18 & 3.4 $\times$ 10$^{-11}$ & 5.3 $\times$ 10$^{-10}$ \\
0.6 & 5.26 & 9.30 & 7.4 $\times$ 10$^{-11}$ & 1.1 $\times$ 10$^{-9}$ \\
0.7 & 6.14 & 8.43 & 1.7 $\times$ 10$^{-10}$ & 2.6 $\times$ 10$^{-9}$ 
\enddata
\tablenotetext{a}{The value quoted is the lower limit to the accretion rate, since no bolometric correction is made to the derived accretion luminosity.  The error on the flux measurement (and therefore the accretion rate) due to uncertainties in the extinction law towards RS Oph is $\sim$60\%.  The mass accretion rate values assume a distance of 1.6 kpc.}
\end{deluxetable}

\section{Spectral Analysis}
\label{spectra}
\subsection{\textit{Chandra}}
The high quality of the \textit{Chandra} data, presented in Figure 4, reveals a number of notable spectral features that can guide us in developing a model of the emission.  We observe soft emission, peaking around 1 keV, comprised of a number of spectral lines.  The most obvious of these can be identified as the unresolved He-like line triplets of Mg XI, Si XIII and S XV at $\sim$1.3, 1.8 and 2.4 keV, respectively.  We also observe an emission feature at $\sim$6.5 keV, which is likely the unresolved blend of some combination of the neutral iron fluorescence line at 6.4 keV, the He-like Fe triplet at 6.7 keV, and the Lyman $\alpha$ line of H-like Fe at 6.9 keV.  

Finally, there is a large, sharp decrease in continuum flux at energies higher than 7 keV.  This absorption feature is due to the 7.1 keV K edge of iron, and is an unmistakable signature of strong absorption of the hard X-ray continuum.  We can estimate the column density of Fe by comparing the count rate on either side of the edge (equivalent to flux since the instrumental effective area is approximately constant over this energy range) and using the following relationship:

\begin{equation}
N(Fe) = -ln(F_{out}/F_{in})/\sigma_{FeK}
\end{equation} where $F_{out}$ is the flux after the edge, $F_{in}$ the flux before the edge (excluding the emission line feature), and $\sigma_{FeK}$ the cross section of the Fe K edge.  Following \citet{Leahy93}, we adopt a $\sigma_{FeK}$ of 3.6 $\times$ 10$^{-20}$ cm$^{2}$. In the \textit{Chandra} spectrum, the flux drops by a factor of $\sim$5 at the Fe K edge, giving N(Fe) = $\sim$4 $\times$ 10$^{19}$ cm$^{-2}$.  This can be converted to a hydrogen column, N(H), by multiplying by the ratio of the numbers of H and Fe atoms in a solar abundance gas.  Using the abundances of \citet{Grevesse98}, the observed Fe edge implies a column density N(H) $\sim$ 1.4 $\times$ 10$^{24}$ cm$^{-2}$.  High column density absorption has been observed in other symbiotic stars, including RT Cru \citep{Luna07} and T CrB, another recurrent nova \citep{Luna08}.  Since such a high column density absorber removes nearly all flux below $\sim$ 2 keV (indeed, the spectra of both RT Cru and T CrB have almost no flux at low energies), we should expect any model of the emission in RS Oph to be comprised of multiple components subject to different absorption in order to simultaneously account for the large soft X-ray flux observed below 1 keV, and the deep Fe K edge.  

\begin{figure}
\begin{center}
\includegraphics[angle=270,width=3.25in]{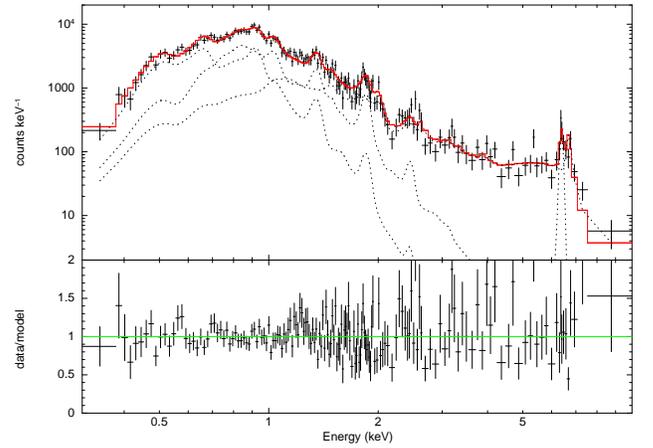}
\caption{\textit{Chandra} ACIS-S spectrum with model CA-va. The unbinned data have been smoothed for the purposes of plotting to give a minimum signal to noise ratio of 3 in each bin. Dotted lines indicate the individual components of the model, and the ratio of the model to the data is shown in the lower panel.}
\end{center}
\end{figure}

We developed a model for the \textit{Chandra} data using Xspec v12.0.5 \citep{Arnaud96} and the binned spectrum in the energy range 0.3--10.0 keV.  All quoted uncertainties are the 90\% confidence intervals obtained using the {\tt error} command in Xspec.  If during the execution of this command a better fit was obtained (based on the $\chi^{2}$ statistic), the error analysis was restarted using the new best fit parameters.  This process was repeated, iterating on the parameter values, until a stable model was found and the error command completed without finding a new fit.  Throughout this section, we quote distance dependent quantities in units of $D$ = 1.6 kpc \citep{Hjellming86}, which was originally derived from HI 21 cm radio data.  This distance is the one most commonly used in the literature, and is in agreement with the value 1.4$^{+0.6}_{-0.2}$ kpc derived by \cite{Barry08} from a survey of all available distance estimate methods.  We will discuss the implications of using larger distances in Section 6.3.

We used the {\tt apec} models in Xspec, developed by \citet{Smith01}, which model both continuum and line emission from thermal plasmas.  We found no acceptable fits for solar abundance plasma models, even with up to three temperature components subject to different absorbing columns.  This is not surprising, since there is clear evidence of non-solar abundances in the RS Oph system from both outburst and quiescence observations.  Based on analysis of high resolution grating spectra of RS Oph obtained during the 2006 outburst, \citet{Nelson08a} and \citet{Ness09} both concluded that the X-ray emitting material in RS Oph is enhanced in nitrogen.  Furthermore, \citet{Pavlenko08} concluded from near-infrared spectroscopy of CO and CN molecular features that the wind from the red giant is depleted in carbon and enhanced in nitrogen, possibly due to either CNO processing in the hydrogen-burning shell of the red giant or contamination from earlier nova outbursts.  

Therefore, in order to explore the effects of non-solar abundances on our model fits, we make use of the values determined by Ness et al. (2009) from the June 2006 \textit{Chandra} LETG observation of RS Oph.  This observation occurred after the end of the supersoft source stage, when the X-rays likely originated in the shocked ejecta, and therefore are the most appropriate to compare with the quiescence data.  This set of abundances, taken from Table 7 in Ness et al., is presented in the first column of Table 3.  The values presented are relative to the solar values of \citet{Grevesse98}, and we assume solar values for elements not listed in Table 3.  It is important to note that the Ness et al. abundances were determined using an emission measure distribution method that assumed a solar abundance of oxygen, which may or may not be appropriate for emission from the nova ejecta.  We implemented these abundances in Xspec using the {\tt vapec} model, which allows the abundance of each element to be varied individually.

\begin{deluxetable}{ccccc}
\tabletypesize{\footnotesize}
\tablecaption{Elemental abundances used in model components}
\tablewidth{0pt}
\tablehead{
 & \multicolumn{4}{c}{Value relative to solar\tablenotemark{a}} \\
 \cline{2-5} \\
\colhead{Element} & \colhead{N09} & \colhead{{\tt vapec}} & \colhead{{\tt vmcflow}} & \colhead{average}
}
\startdata
O & 1.00\tablenotemark{b} & 0.7$\pm$0.1 & 0.7$^{+0.2}_{-0.1}$ & 0.7$^{+0.2}_{-0.1}$ \\
N & 7.54$^{+12}_{-5}$ & 7.54 & 7.54 & 7.54\\
Ne & 1.36 $\pm$ 0.16 & 1.36 &1.36 & 1.36\\
Mg & 1.91 $\pm$ 0.70 & 1.5 $\pm$ 0.3 & 1.6$^{+0.4}_{-0.3}$ & 1.55$^{+0.4}_{-0.3}$\\
Si & 1.62 $\pm$ 1.00 & 1.62 & 1.62 & 1.62\\
Fe & 0.43 $\pm$ 0.10 & 0.6 $\pm$ 0.1 & 0.6$^{+0.2}_{-0.1}$ & 0.6$^{+0.2}_{-0.1}$
\enddata
\tablenotetext{a}{Solar abundances are taken from Grevesse and Sauval (1998).}
\tablenotetext{b}{These abundances were derived using the Emission Measure Distribution (EMD) method described in Ness at al. (2009), which assumes a solar abundance of oxygen.}
\end{deluxetable}

We find no acceptable fit ($\chi^{2}$/$\nu$ = 2.9 for $\nu$ = 101) for a two plasma model, where each component is subject to differing absorption.   Adding a third absorbed plasma component improves the fit, particularly at the soft end of the spectrum, with $\chi^{2}$/$\nu$ = 2.1 for $\nu$ = 98, although it is still not statistically acceptable.  The soft flux arises in two components with kT = 0.2 and 0.6 keV, subject to absorbing columns of 4.7 and 1.0 $\times$ 10$^{21}$ cm$^{-2}$, respectively.  These values of N(H) are about a factor of $\sim$2 higher and lower, respectively, than the interstellar value of 2.4 $\pm$ 0.6 $\times$ 10$^{21}$ cm$^{-2}$ reported by \citet{Hjellming86} from HI 21cm radio observations. The flux above 2 keV originates in a third component, whose temperature in this model is 69 keV.  However, the fit to the data (particularly above 5 keV) is extremely poor, and the model has a much steeper slope at these energies than is observed in the data.  Also, the column density of the absorber attenuating this component is extremely small (3.7 $\times$ 10$^{19}$ cm$^{-2}$), which is completely at odds with the observed Fe K edge at 7.1 keV.  The low column density of this absorber is required by the model in order to reproduce the continuum slope between 2 and 5 keV - a complete absorber with a column density of $\sim$10$^{24}$ cm$^{-2}$ removes all 0.3--2 keV, and results in a rising continuum energies greater than 2 keV, which is clearly not observed in the data.  This discrepancy, considered in tandem with best fit N(H) values which differ significantly from the  interstellar value towards RS Oph, made us search for alternative models which can account for the spectral properties in a more consistent way. 

A possible answer to the question of the physical origin of the X-ray emission of RS Oph may lie in X-ray studies of other symbiotic stars.  The fits to the X-ray spectra of RT Cru and T CrB both require the presence of an absorber that only partially covers the X-ray emitting source to explain their 2--10 keV emission.  Although difficult to disentangle in RS Oph (where there is clearly an additional, less absorbed soft component) we also tried to fit the continuum emission with a model comprised of a completely absorbed soft component, and a partially covered, absorbed hard component.  This has the effect of allowing soft flux from the hard component to contribute to the spectrum, even though the column in front of the hard X-ray emitting material is very high.  We used the {\tt pcfabs} model in Xspec for this case, which is characterized by an absorbing column N(H)$_{pc}$, and a covering fraction (C. F. $\leq$ 1) that quantifies the degree to which the absorber covers the source.

The fit statistic for this model (which we implemented as {\tt wabs*vapec+pcfabs*vapec} in Xspec) is much better than the model with a completely covered hard component, with $\chi^{2}$/$\nu$ = 1.5 for $\nu$ = 119.  The temperature of the soft component is 0.35$^{+0.07}_{-0.04}$ keV subject to an absorbing column N(H)$_{tot}$ = 2.4$^{+0.2}_{-0.4}$ $\times$ 10$^{21}$ cm$^{-2}$, which is consistent with the ISM contribution.  The hard component has a temperature of 3.1$^{+0.9}_{-0.5}$ keV.  The absorber has a column density N(H)$_{pc}$ = 1.3 $\pm$ 0.2  $\times$ 10$^{24}$ cm$^{-2}$, and covers 98 $\pm$ 1 \% of the source.  There is a narrow, 3$\sigma$ residual feature at approximately 6.4 keV, which likely indicates the presence of the Fe K alpha fluorescence line.  This line is produced via fluorescence after ionization of a K shell electron.  Given the high optical depth of the Fe K edge, the K $\alpha$ line should be bright.  We therefore added a gaussian to the model at 6.4 keV, and kept the width consistent with instrumental broadening only.  The best fit energy of the line is 6.4 $\pm$ 0.04 keV, and the equivalent width is 320$^{+480}_{-270}$ eV.  All other quantities related to the harder plasma and absorber remain the same within the uncertainties as the model with no gaussian.  

Examining the residuals of the model fit to the data, we see that the model predicts more flux than is observed at energies below 0.5 keV.  All of this excess flux arises in the partially covered component of the model.  However, even this component should be subject to interstellar absorption along the line of sight.  Since the absorption affecting the soft component is consistent with being due only to the ISM, we altered the model slightly so that all components are affected by a complete absorber, representing the ISM, and the hard component further attenuated by the partial covering absorber.  In Xspec, this was implemented as {\tt wabs(vapec+pcfabs(vapec+gauss))}.  Although this modification makes physical sense, it does not formally improve the fit, giving a $\chi^{2}$/$\nu$ = 1.5 for $\nu$ = 100.  In fact, the model now underpredicts the flux at energies less than 0.6 keV.  As a final refinement, we added an additional plasma absorbed only by the total absorber, implemented as {\tt wabs(vapec+vapec+pcfabs(vapec+gauss))} in Xspec.  This results in an improved reduced $\chi^{2}$ value of 1.3 for $\nu$ = 98.  The first two plasma components have temperatures of 0.14$^{+0.03}_{-0.02}$ and 0.35$^{+0.02}_{-0.01}$ keV, absorbed by a column with N(H)$_{tot}$ = 2.1$^{+0.3}_{-0.4}$ $\times$ 10$^{21}$ cm$^{-2}$, which is still consistent with previous estimates of the ISM contribution towards RS Oph.  The third plasma component has kT = 2.5 $^{+0.5}_{-0.3}$ keV, and the partial covering absorber has N(H)$_{pc}$ = 1.1$^{+0.3}_{-0.2}$ $\times$ 10$^{24}$ cm$^{-2}$, with a 98 $\pm$ 1 \% covering fraction.  

We note that we have binned the data in order to utilize the $\chi^{2}$ statistic in developing our model.  Such binning unavoidably leads to a loss of spectral information.  In order to check if our parameter values were in anyway influenced by our choice of binning, we refit our model using the unbinned data and the Cash statistic \citep{Cash79} as implemented in XSpec.  We find no difference in the best fit model parameters using the different statistic, and the range of uncertainties is also similar, showing that our model fitting is not influenced in any strong way by the choice of binning.  However, since the fit using the Cash statistic makes use of all the available spectral information, we show the parameters and uncertainties derived using that statistic in Table 4, as model CA.  We plot the data with this model in Figure 4, and have smoothed the data for plotting purposes to give a minimum signal to noise ratio in each bin of 3.
\begin{deluxetable*}{lcccc | cccc}
\tabletypesize{\footnotesize}
\tablecaption{Model fits to the \textit{Chandra} spectra}
\tablewidth{0pt}
\tablehead{
 & \multicolumn{4}{c}{\textbf{Complete observation}} & \multicolumn{2}{c}{\textbf{First half} }& \multicolumn{2}{c}{\textbf{Second half}} \\
\colhead{Parameter} & \colhead{CA} & \colhead{CB} & \colhead{CA-va} & \colhead{CB-va} &
\colhead{C1A} & \colhead{C1B} & \colhead{C2A} & \colhead{C2B} 
}
\startdata
N(H)$_{total}$ (10$^{21}$ cm$^{-2}$)                                          & 1.9$^{+0.4}_{-0.4}$       & 1.8$^{+0.3}_{-0.4}$       & 2.1$\pm$ 0.5                  & 2.1$^{+0.6}_{-0.5}$       & 2                                    & 2                                    & 2                                      & 2\\
kT$_{1}$ (keV)                                                                               & 0.15$^{+0.06}_{-0.03}$ & 0.16$^{+0.08}_{-0.04}$ & 0.25$^{+0.04}_{-0.02}$ & 0.25$^{+0.02}_{-0.03}$ & 0.26$^{+0.02}_{-0.05}$ & 0.26$^{+0.03}_{-0.07}$ & 0.25$^{+0.02}_{-0.03}$  & 0.25$\pm$0.03  \\
norm$_{1}$ (10$^{-4}$)                                                                  & 0.4$^{+0.5}_{-0.3}$       & 0.3$^{+0.5}_{-0.2}$       & 2.1$^{+1.2}_{-0.7}$       & 2.1$^{+1.3}_{-0.7}$       & 2.2$^{+2.9}_{-1.7}$       & 2.2$^{+3.3}_{-1.1}$       & 1.9$\pm$0.3                   & 1.9$\pm$0.3 \\
kT$_{2}$  (keV)                                                                              & 0.37$^{+0.02}_{-0.01}$  & 0.37$^{+0.03}_{-0.01}$ & 0.51$^{+0.10}_{-0.06}$ & 0.53$^{+0.09}_{-0.08}$ & 0.43$\pm$0.13              & 0.43$^{+0.13}_{-0.25}$ & 0.60$\pm$0.04               & 0.60$^{+0.04}_{-0.06}$ \\
norm$_{2}$ (10$^{-4}$)                                                                  & 2.6$^{+0.4}_{-0.5}$       & 2.3$^{+0.5}_{-0.4}$       & 1.3$^{+0.5}_{-0.6}$       & 1.1$^{+0.5}_{-0.4}$       & 0.9$^{+1.8}_{-0.5}$       & 0.9$^{+1.6}_{-0.7}$       & 1.5$^{+0.4}_{-0.3}$       & 1.5$^{+0.3}_{-0.5}$  \\
N(H)$_{pc}$\tablenotemark{a} (10$^{22}$ cm$^{-2}$)                 & 117$^{+31}_{-25}$         & 120$^{+40}_{-22}$        & 106$^{+30}_{-23}$        & 108$^{+38}_{-26}$        & 84$^{+30}_{-19}$          & 86$^{+20}_{-23}$          & 113$^{+35}_{-36}$        & 118$^{+60}_{-40}$ \\
C. F.\tablenotemark{b}                                                                   & 0.98$\pm$ 0.01              & 0.98$\pm$0.01                & 0.98$\pm$ 0.01             & 0.97$\pm$0.02              & 0.98$^{+0.01}_{-0.02}$ & 0.97$^{+0.01}_{-0.02}$ & 0.97$^{+0.01}_{-0.03}$ & 0.97$^{+0.02}_{-0.04}$\\
kT$_{3}$ (keV)		                                                                      & 2.5$^{+0.5}_{-0.3}$        & \nodata                           & 2.6$^{+0.7}_{-0.4}$      & \nodata                          & 2.4$^{+0.8}_{-0.5}$       & \nodata                           & 2.6$^{+0.7}_{-0.5}$       & \nodata \\
norm$_{3}$  (10$^{-3}$)                                                                 & 7$^{+6}_{-3}$                 & \nodata                           & 5$^{+4}_{-2}$               & \nodata                          & 4$^{+5}_{-2}$                & \nodata                           & 5$^{+6}_{-3}$                & \nodata \\
kT$_{min}$ (keV)                                                                            & \nodata                           &$<$1.8                            & \nodata                          & $<$1.2                            & \nodata                          & 1.6$^{+1.7}_{-1.5}$        & \nodata                          & $<$3.7 \\
kT$_{max}$  (keV)                                                                          & \nodata                           & 5.4$^{+2.4}_{-1.1}$       & \nodata                         & 5.9$^{+3.5}_{-1.6}$        & \nodata                          & 3.7$^{+14.4}_{-1.7}$      & \nodata                          & 4.2$^{+4.8}_{-1.7}$ \\
$\dot{M}$/$D^{2}_{1.6 kpc}$ (10$^{-9}$ M$_{\odot}$ yr$^{-1}$)   & \nodata                          & 2.6$^{+2.7}_{-1.9}$        & \nodata                         & 1.6$^{+1.9}_{-0.9}$        & \nodata                          & 3.6$^{+46.4}_{-1.9}$      & \nodata                          & 3.8$^{+110.9}_{-3.2}$\\
E$_{gauss}$ (keV)                                                                          & 6.39$^{+0.03}_{-0.04}$ & 6.39$^{+0.03}_{-0.04}$ & 6.39$^{+0.03}_{-0.04}$ & 6.39$\pm$0.03               & 6.39                               & 6.39                                & 6.39                               & 6.39 \\
norm$_{gauss}$ (10$^{-6}$)                                                           & 11$^{+9}_{-5}$              & 11$^{+11}_{-5}$            & 9.3$^{+6.9}_{-4.3}$       & 9.4$^{+8.8}_{-3.8}$       & 12.2$^{9.7}_{-5.0}$       & 11$^{+11}_{-5}$             & 4.7$^{+9.2}_{-4.7}$       & 4.9$^{+8.0}_{-3.8}$ \\
\cline{1-9}
Fe                                                                                                     & 0.43                               & 0.43                               & 0.6$\pm$ 0.1                 & 0.6$^{+0.2}_{-0.1}$       & 0.6                                  & 0.6                                  & 0.6                                 & 0.6 \\
O                                                                                                       & 1.0                                & 1.0                                 & 0.7$\pm$0.1                  & 0.7$^{+0.2}_{-0.1}$       & 0.7                                  & 0.7                                   & 0.7                                & 0.7 \\
Mg                                                                                                     & 1.91                              & 1.91                               & 1.5$\pm$ 0.3                 & 1.6$^{+0.4}_{-0.3}$       & 1.55                                & 1.55                                 & 1.55                              & 1.55 \\
\cline{1-9} 
\cline{1-9}
$\chi^{2}$/$\nu$ ($\nu$)                                                                    & 1.27 (96)                       &  1.29 (95)                       & 1.03 (93)                       & 1.04 (92)                       & 0.90 (53)                        &  0.93 (52)                      & 0.84 (75)                          & 1.04 (92) \\
C-statistic ($\nu$)                                                                              & 773 (652)                 & 775 (651)                  & 742 (649)                & 743 (648)                  & 608 (654)                 &  610 (653)                 & 619 (654)                    & 621 (653)

\enddata

\tablecomments{"A" models are of the form {\tt wabs(vapec+vapec+pcfabs(vapec+gauss))}, i.e. have a single temperature plasma as the hard component.  "B"  models are of the form {\tt wabs(vapec+vapec+pcfabs(vmcflow+gauss))} i.e. have a cooling flow as the hard component.  Models CA and CB assume the abundances of Ness et al. 2009, as outlined in Section 4.  In models CA-va and CB-va, the abundances of Fe , O and Mg were allowed to vary freely (see Section 4.1.2).  The separate fits to the first (C1 models) and second half (C2 models) of the observation use the average abundance values found in models CA-va and CB-va.  All quoted uncertainties are 90\% confidence intervals obtained using the {\tt error} command in Xspec.  Parameters with no uncertainties were not allowed to vary, as discussed in the body of the text.}
\tablenotetext{a}{The column density of the partial covering absorber.}
\tablenotetext{a}{The covering fraction of the partial covering absorber.}
\end{deluxetable*}

\subsubsection{A cooling flow origin for the hard component}
The X-ray spectra of many accreting white dwarfs, including the hard X-ray detected symbiotic stars, can be well described by a cooling flow model \citep{Mukai03}. In this model, which was originally developed for the gas in galaxy clusters, hot material in the post shock region of the accretion flow cools isobarically and settles onto the white dwarf surface (see \citet{Fabian94} for a review of the cooling flow model).  The model assumes no additional sources of heating or cooling once the accreted material is in the boundary layer.  While this is generally the case for non-magnetic white dwarfs (for magnetic systems, cyclotron cooling can be important),  it may not in fact be true for RS Oph, as we will discuss in Section 6.  The model is particularly useful for studying accretion in CVs, since it provides an estimate of the accretion rate onto the white dwarf directly from model fitting.  To determine whether the RS Oph data can be described by this model, we replaced the partially absorbed {\tt vapec} component in model A with {\tt vmcflow}, the variable abundance implementation of the cooling flow model in Xspec.  We used the Ness et al. (2009) abundances, and allowed the maximum and minimum temperatures of the cooling flow to vary freely.  

The resulting model has an equally good fit to the data as the {\tt vapec} model ($\chi^{2}$/$\nu$ = 1.29, $\nu$ = 95). We checked the fit result using the unbinned data and the Cash statistic, and again found no differences in best fit parameter values within the uncertainties.  We present the best fit parameters derived using the Cash statistic for this model (model CB) in Table 4.  The parameters for the soft component (i.e the two {\tt vapec} components) agree with those of model CA within the uncertainties.  The parameters of the total and partial covering absorbers are also the same within the uncertainties.  We find minimum and maximum temperatures for the cooling flow of $<$0.97 keV and 5.9$^{+1.8}_{-1.2}$ keV, respectively. The upper limit to the minimum temperature indicates that cooling within the flow is efficient.  The maximum temperature in the cooling flow is surprisingly low, and we will discuss this result further in Section 6.2.  Finally, the normalization gives an estimate of the cooling flow accretion rate, in this case 2.4$^{+2.1}_{-1.2}$ $\times$ 10$^{-9}$ ($D$/1.6 kpc)$^{2}$ M$_{\odot}$ yr$^{-1}$, where $D$ is the distance to the source in kpc.

\subsubsection{Effects of different abundance?}

We explored whether allowing the abundances of our two best fit models to vary would improve the fit.  This was motivated primarily by the fact that the abundance estimates of Ness et al. (2009) have an associated uncertainty.  Determining abundances from CCD resolution spectrum is difficult, since few spectral features are available for use in model fitting.  In this case, we allowed the abundances of Ne, Mg, Si and Fe to vary freely, since these spectral features are clearly visible in the data.  We also explored the effect of changing the O abundance, since a clear residual at 0.6 keV (due to O VIII Ly $\alpha$) is observed in the fit using the Ness abundances (which were determined assuming solar oxygen, see Section 4.1).  For each of these elements, we identified the model component contributing the majority of the flux, and allowed the elemental abundance parameter in that component to vary freely.  The associated abundance parameters in the other model components were tied to this value.  By doing this, all spectral components varied in tandem, but the model fitting was guided by the component in which the spectral feature of interest was strongest.

We initially varied each elemental abundance individually, starting from either model CA or CB, and used the F--test to determine whether changing the abundance parameter resulted in a statistically significant improvement in the reduced $\chi^{2}$ value.  We find that allowing Fe, O or Mg to vary freely results in an improved fit that is significant at a level greater than 99.9\%.  The change in the Fe abundance is the most significant in terms of the F-test result.  Freeing the Ne abundance improves the fit marginally, but the F--test null hypothesis can only be rejected at the 75\% level, so the improvement is probably not significant.  Varying Si results in no change to the fit statistic.  We find similar improvements in the fit to both the single temperature and cooling flow models developed above.

Next, we refit the models allowing the O, Mg and Fe abundances to all vary freely.  The resulting models have smaller $\chi^{2}$/$\nu$ values and residuals when compared to the Ness et al. (2009) abundances.  Once again, fitting the unbinned data using the Cash statistic did not change the best fit parameters within the uncertainties determined using the $\chi^{2}$ statistic.  We present the resulting abundances obtained for each model (and their average values) in Table 3, and the model parameters (model CA-va with {\tt vapec} hard component; model CB-va with {\tt vmcflow} hard component, where ``-va" indicates that abundances were varied) obtained using the C-statistic method, in Table 4. The best fit values agree within uncertainties between the two models.  We actually determine a subsolar abundance of O, in contrast with the assumption of solar O in the DEM method of Ness et al..  However, we note that our best fit value (0.7$^{+0.2}_{-0.1}$) is in good agreement with the O abundance determined by Ness et al. from fitting {\tt apec} models to the Day 26 grating data, a method that estimates absolute abundance corrections by comparing line strengths with the continuum.  We find a higher Fe and lower Mg abundance than Ness et al. (0.6$^{+0.2}_{-0.1}$ and 1.55$^{+0.4}_{-0.3}$, respectively), although within the uncertainties they are in agreement. The best fit temperatures and normalizations of the two soft plasmas are different than in models CA and CB (see Table 4), but the hard component parameters remain in agreement within the uncertainties.  

\subsubsection{Count rate and spectral changes during the \textit{Chandra} observation}

\begin{figure}
\begin{center}
\includegraphics[angle=270,width=3.25in]{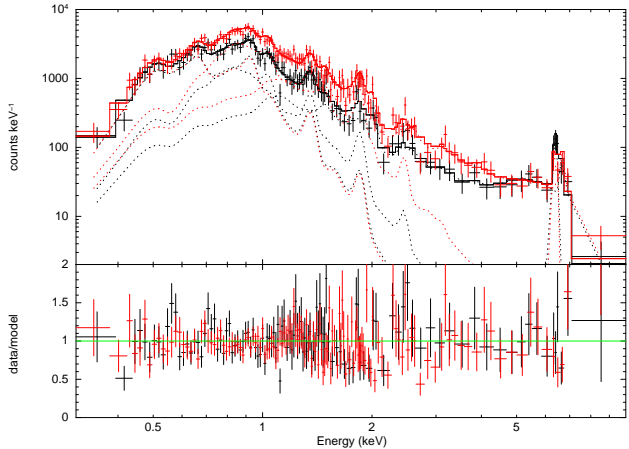}
\caption{Comparison of the spectra obtained from the first half (\textit{black data points}) and second half (\textit{red data points}) of the \textit{Chandra} observation.  We also show the associated models C1A and C2A. The data have been smoothed as in Figure 4.  The ratio of the model to the data for each spectrum is shown in the lower panel.}
\end{center}
\end{figure}

In order to explore the physical origin of the count rate variations observed in the Chandra lightcurve, we extracted two separate spectra from the observation---one from the first half, where the count rate was increasing, and one from the second half where it was approximately constant.  The time dividing the observation is marked in Fig 1 by the dashed vertical line. We used the best fit models found for the total spectrum as the starting point for fitting the split spectra.  Since looking at smaller time periods reduces the signal to noise of the spectrum, we have fixed the values of a number of parameters in order to reduce the number of degrees of freedom.  The N(H) of the total absorber  (N(H)$_{tot}$) is consistent with the ISM contribution in all models, and therefore we set N(H)$_{total}$ =  2 $\times$ 10$^{21}$ cm$^{-2}$.  We set the Fe, O and Mg abundances to the average values from models C and D i.e. 0.6, 0.7 and 1.55 times the solar values.  Finally, we set E$_{gauss}$ to 6.39 keV.  Reducing the degrees of freedom in this reasonable way makes a stable fit more likely, and allows us to explore which parameters are responsible for the change in count rate.  The resulting best fit parameters and associated uncertainties are listed in the second half of Table 4.  The spectra and models C1A and C2A are presented in Figure 5, with the first half of the observation in black, and the second in red.  Overall, the two spectra are quite similar---within the uncertainties in each bin, they are the same below 0.8 keV.  They are also similar above 2 keV, although the Fe K$\alpha$ feature at 6.4 keV is brighter in the first half than in the second.  The main differences between the two spectra occur between 0.8 and 2.0 keV, as the lightcurves indicated.

However, we cannot definitively identify the origin of this change from our best fit models---within the formal statistical errors there is no difference between the two halves of the spectrum.  Comparing the two halves using the $\chi^{2}$-statistic, the parameters of the soft plasmas agree within uncertainties, and all changes appear to be in the hard component (both in the column density of the absorber, and in the normalization of the hard plasma).  However, checking the fits using the unbinned data and the C-statistic, we find instead that the change in 1-2 keV count rate is due to a change of temperature and normalization in the second soft plasma component, with no detectable difference in the hard plasma.

\subsection{\textit{XMM-Newton}}
The high background level during parts of the \textit{XMM-Newton} observation resulted in a net exposure time of only $\sim$21 ks for the pn camera, and $\sim$28 ks for the MOS cameras.  Thus there are only a small number of counts above 2 keV, which increases the challenge of determining the best fit model to the data.  We fit our model to the spectra obtained with all three cameras simultaneously, and we used the models developed for the \textit{Chandra} data in Table 4 as the starting parameter values.  As we did for the split spectra discussed in Section 4.1.3, we fixed the N(H)$_{tot}$ (the column density of the total absorber) to 2 $\times$ 10$^{21}$ cm$^{-2}$, and the energy of the gaussian component to 6.39 keV to decrease the number of degrees of freedom.  Again, since the total absorber is consistent with the ISM contribution, which should not evolve with time, we are justified in constraining this parameter.  We used two sets of abundances - those of Ness et al. (2009), and the updated and averaged Fe, O and Mg abundances we found from the \textit{Chandra} observation,  to see if there were differences in the fit parameters, as we found with the \textit{Chandra} data.

\begin{deluxetable*}{lcccc}
\tabletypesize{\footnotesize}
\tablecaption{Model fits to the \textit{XMM-Newton} spectrum}
\tablewidth{0pt}
\tablehead{
\colhead{Parameter} & \colhead{Model XA} & \colhead{Model XB} & \colhead{Model XA-n} & \colhead{Model XB-n} 
}
\startdata
N(H)$_{total}$ (10 $^{21}$ cm$^{-2}$)                                            & 2                                         & 2                                           & 2                                         & 2 \\
kT$_{1}$ (keV)                                                                                      & 0.12$^{+0.03}_{-0.02}$ & 0.12$^{+0.02}_{-0.03}$   & 0.13$^{+0.04}_{-0.03}$  & 0.13$^{+0.04}_{-0.03}$   \\
norm$_{1}$ (10$^{-4}$)                                                                      & 1.1$^{+0.5}_{-0.4}$        & 1.1$^{+1.0}_{-0.5}$          & 0.9$^{+0.5}_{-0.3}$          & 0.8$^{+0.8}_{-0.3}$  \\
kT$_{2}$  (keV)                                                                                     & 0.31 $\pm$ 0.01             & 0.31 $\pm$ 0.02                & 0.30$^{+0.02}_{-0.01}$  & 0.29$^{+0.02}_{-0.01}$  \\
norm$_{2}$ (10$^{-4}$)                                                                      & 1.7$^{+0.1}_{-0.2}$        & 1.7$^{+0.1}_{-0.2}$          & 1.9$^{+0.2}_{-0.3}$         & 1.9$^{+0.2}_{-0.3}$  \\
N(H)$_{pc}$\tablenotemark{a} (10$^{22}$ cm$^{-2}$)                & 111$^{+36}_{-58}$        & 109$^{+54}_{-44}$             & 124$^{+73}_{-45}$        & 119$^{+72}_{-43}$  \\
C. F.\tablenotemark{b}                                                                         & 0.96$^{+0.03}_{-0.04}$ & 0.95$^{+0.03}_{-0.03}$   & 0.95$^{+0.03}_{-0.06}$  & 0.95$^{+0.03}_{-0.05}$ \\
kT$_{3}$ (keV)		                                                                            & 6$^{+9}_{-3}$                  &   \nodata                              & 10$^{+34}_{-4}$                               & \nodata \\
norm$_{3}$  (10$^{-3}$)                                                                      &  0.6$^{+1.0}_{-0.4}$       & \nodata                               & 0.6$^{+1.0}_{-0.3}$                            & \nodata \\
kT$_{min}$ (keV)                                                                                  & \nodata                              & $<$11                                & \nodata                               & $<$ 4.4\\
kT$_{max}$  (keV)                                                                                & \nodata                             & 14$^{+47}_{-8}$                               &  \nodata                              & $>$12 \\
$\dot{M}$/$D^{2}_{1.6 kpc}$ (10$^{-9}$ M$_{\odot}$ yr$^{-1}$) &  \nodata                            & 0.1$^{+21}_{-0.07}$     & \nodata                                & 0.09$^{+0.22}_{-0.07}$ \\
E$_{gauss}$ (keV)                                                                               & 6.39                                   & 6.39                                    & 6.39                                     & 6.39 \\
norm$_{gauss}$                                                                                    & 2.3$^{+7.2}_{-1.8}$        & 2.2$^{+6.9}_{-1.7}$        & 2.4$^{+6.7}_{-2.0}$          & 2.2$^{+5.8}_{-1.9}$  \\
\cline{1-5}
Fe                                                                                                             & 0.43                                    & 0.43                                     & 0.6                                      & 0.6 \\
O                                                                                                               & 1.0                                      & 1.0                                       & 0.7                                      & 0.7 \\
Mg                                                                                                           & 1.91                                    & 1.91                                     & 1.55                                     & 1.55 \\
\cline{1-5}
\cline{1-5}
$\chi^{2}$/$\nu$ ($\nu$)                                                                       & 1.61 (85)                              &  1.64 (84)                           & 1.42 (85)                            & 1.44 (84) \\
C-statistic ($\nu$)                                                                                   & 2732 (5811)                      & 2732 (5810)                     & 2734 (5811)                     & 2732 (5810)

\enddata
\tablecomments{The form of these models is the same as that outlined in Table 4.  Again, parameters with no uncertainties were fixed as outlined in the text.  The ``-n" models denote that the abundances determined in this paper were used.}

\tablenotetext{a}{The column density of the partial covering absorber.}
\tablenotetext{b}{The covering fraction of the partial covering absorber.}

\end{deluxetable*}

The resulting best fits to the \textit{XMM-Newton} data are presented in Table 6 (models XA, XB, XA-n and XB-n), and the data and model XA-va are presented in Figure 6.  Here, the ``n" models use the average new abundances found from the fits to the {\it Chandra} spectrum in this paper, like the fits to the partial spectra in Table 4.   The emission below 2 keV is still well described by a two temperature plasma model with kT values between 0.14 and 0.62 keV.  In contrast with the \textit{Chandra} analysis, there is no clear difference in the soft plasma best fit parameters between the two sets of abundances.  The temperature values agree with those found in the \textit{Chandra} observation within the uncertainties.  However, the normalization of these components (and thus the emission measure of the plasmas) has decreased.  While a hard component is still clearly seen in the data, the parameters are poorly constrained due to the small number of counts.   The best fit values for the normalization (single temperature) or accretion rate (cooling flow) of the hard component are much lower than for the Chandra observation, although the formal confidence intervals are large and do not completely rule out higher values.  This large uncertainty is driven by degeneracies in the N(H)$_{pc}$, covering fraction and plasma normalization space.  With so few hard counts, there are few features in the data that can eliminate these degeneracies, as we were able to do in the \textit{Chandra} data.  However, the lower observed flux and smaller hard component normalizations suggest that the accretion rate is lower than in the \textit{Chandra} observation.

\begin{figure}
\begin{center}
\includegraphics[angle=270,width=3.25in]{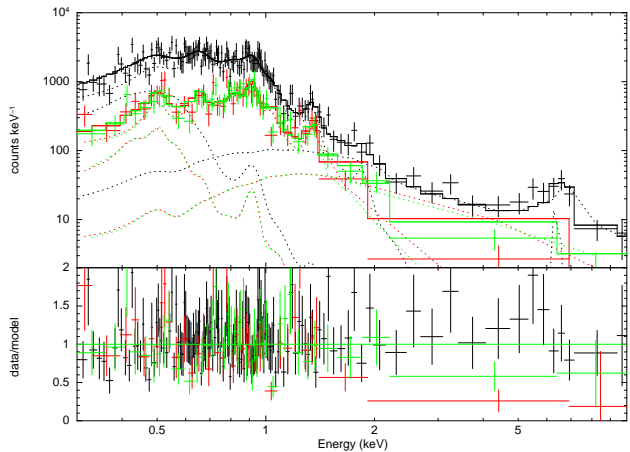}
\caption{\textit{XMM-Newton} EPIC-pn spectrum with model XA-n. The pn data are in black, and the MOS1 and MOS2 data in red and green, respectively.  We show the ratio of the model to the data in the lower panel.  The data have been smoothed as in Figure 4.}
\end{center}
\end{figure}

\subsection{Constraints on an optically thick boundary layer}
Several authors have shown that at high accretion rates, the boundary layers in CVs should become optically thick, thermalize, and emit like a blackbody with effective temperature of a few tens of eV (depending on $\dot{M}$); see e.g. \citet{Patterson85}.  This blackbody component would be subject to the same absorption as the hard component.  Therefore, in order to place constraints on the luminosity and temperature of an optically thick boundary layer, we added a blackbody component to our best fit \textit{Chandra} and \textit{XMM-Newton} models.  The luminosity of this optically thick portion of the boundary layer is given by
\begin{equation}
L_{BL} = 4\pi~R_{WD}^{2}~\sigma~T_{eff}^{4}~f_{BL}
\end{equation} where $f_{BL}$ is the fraction of the white dwarf surface covered by the optically thick boundary layer, expected to be $\leq$10\% \citep[see e.g.][]{Popham95,Piro04}.  In our model we assume $R_{WD}$ =  3 $\times$ 10$^{8}$ cm, $f_{BL}$ = 0.1, leave $T_{eff}$ as the free parameter, and tie the normalization constant of the blackbody component (the luminosity) to $T_{eff}$ via the above equation, assuming a distance of 1.6 kpc.  The blackbody component is absorbed by both the interstellar medium and the partial covering absorber in our model.  We find that the maximum kT$_{eff}$ allowed in our model is 0.034 keV ($\sim$395,000 K) for the \textit{Chandra} model, and 0.031 keV ($\sim$360,000 K) for \textit{XMM-Newton}.  For temperatures higher than this, a significant excess begins to emerge over the data at energies $<$0.5 keV.  This result depends somewhat on the partial covering fraction of the complex absorber---the higher the covering fraction, the higher allowable temperature.  

The maximum allowable blackbody temperature value of 0.034 keV for the \textit{Chandra} observation corresponds to an intrinsic luminosity of 1.6 $\times$ 10$^{35}$ ($D$/1.6 kpc)$^{2}$ erg s$^{-1}$.  If this was the only source of emission from the boundary layer, and assuming $M_{WD}$ = 1.3 M$_{\odot}$, we would obtain a value for $\dot{M}$ of 8.5 $\times$ 10$^{-9}$ ($D$/1.6 kpc)$^{2}$ M$_{\odot}$ yr$^{-1}$.  For the \textit{XMM-Newton} data, the maximum temperature implies an accretion luminosity of 1.1 $\times$ 10$^{35}$ ($D$/1.6 kpc)$^{2}$ erg s$^{-1}$, or an accretion rate of  5.7 $\times$ 10$^{-9}$ ($D$/1.6 kpc)$^{2}$ M$_{\odot}$ yr$^{-1}$.

The derived blackbody temperature limits indicate that most of the flux from any optically thick boundary layer would emerge in the UV.  Therefore, as a consistency check, we predicted the flux in the OM UVM2 filter using one of our best fit EPIC-pn model (model XA) plus the maximum allowable blackbody (31 eV, 51.1 $\times$ 10$^{35}$ erg s$^{-1}$). We did this by creating a one channel spectrum file using the average observed UVM2 count rate, and included this data point in our XSpec model by using response files downloaded from the \textit{XMM-Newton} calibration database.  We find that the UV flux predicted by the model is a factor of a few hundred lower than is actually observed.  Thus, adding the blackbody remains consistent with the OM data.

\section{Comparison with previous X-ray observations}
\label{comparison}
We compared the best fit models and fluxes from the current data to those obtained by \textit{ROSAT} and \textit{ASCA} before the 2006 outburst.  The observed count rates of the old observations are low, and therefore the statistics are poor.  However, by comparing our best fit models and observed count rates at different energy bands, we can investigate the long term behavior of the different emission components observed in the system.  In order to estimate equivalent \textit{ROSAT} and \textit{ASCA} count rates, we folded our best fit \textit{Chandra} and \textit{XMM-Newton} models through the appropriate instrumental responses in XSpec.  These count rates were then compared to the previously published data.

The \textit{ROSAT} observations are useful for looking at the soft component of the emission.  In our best fit models, this arises in warm plasma with temperatures ranging from 0.15 to 0.4 keV.  The first \textit{ROSAT} observation, carried out on 1991 March 2, had an observed count rate in the energy range 0.2--2.4 keV of 3.5 $\times$ 10$^{-3}$ c/s.  This had increased to 1.2 $\pm$ 0.2 $\times$ 10$^{-2}$ c/s one year later, on 1992 March 4.  If the source emission was described by our best fit \textit{Chandra} models, all of which result in similar fluxes, then the resulting \textit{ROSAT} 0.15-4.0 keV count rate would have been 2.1 $\times$ 10$^{-2}$ c/s, a factor of two higher than the brightest \textit{ROSAT} observation.  Our best fit \textit{XMM-Newton} model is in better agreement with the data, differing by only 25\% from the observed count rate in the 1993 observations. However, in 1991 the system was much fainter than during our \textit{XMM-Newton} observation.  

Our best fit \textit{Chandra} and \textit{XMM-Newton} models are both consistent with the 1997 \textit{ASCA}/GIS data in the 2--10 keV energy range, primarily due to the poor counting statistics.  The observed GIS count rate in the 2.0--10.0 keV range is 1.2 $\pm$ 0.8 $\times$ 10$^{-3}$ c/s, compared with 1.4 $\times$ 10$^{-3}$  c/s (model CA, \textit{Chandra}) and 3.2 $\times$ 10$^{-4}$ c/s (model XA, \textit{XMM-Newton}). The \textit{Chandra} models over-estimate the flux in the 0.7--2.0 keV range, as observed with  the \textit{ASCA}/SIS instrument.  However, the \textit{XMM-Newton} models are in much better agreement with the data.  

\section{Interpretation of the data and model fits}
\label{discussion}
\subsection{Evidence of high column intrinsic absorption}
All of our best fit models are comprised of a soft and hard component.  The hard component is equally well described by either a single temperature plasma, or a cooling flow, plus a 6.4 keV neutral Fe emission line obscured by a high column density intrinsic absorber which covers most, but not all, of the hard X-ray emitting region.  Extremely high values for the column density of this complex absorber are implied by the presence of a deep Fe K edge at 7.1 keV, and the values of this parameter in our best fit models are between 0.9 and 1.2 $\times$ 10$^{24}$ cm$^{-2}$.  In all models, the covering fraction is $>$95\%.   Similar high column intrinsic absorbers have been required to fit the X-ray spectra of other symbiotic stars; T CrB, RT Cru, CH Cyg \citep{Kennea09}, and perhaps most interestingly the jet producing system MWC 560, which is viewed almost pole-on \citep{Stute09}.  Since only the hard component is affected by this absorption, the absorbing material must be close to the white dwarf.

\citet{Walder08} carried out hydrodynamical modeling of an accretion-outburst cycle in RS Oph, assuming a mass loss rate from the red giant of 10$^{-7}$ M$_{\odot}$ yr$^{-1}$.  They find that a thick disk forms with a strong spiral distortion, and that the accretion rate onto the white dwarf is 10\% of the mass loss rate from the giant.  The densities in the vicinity of the white dwarf are as high as 10$^{-12.5}$ g cm$^{-3}$, which depending on the line of sight through the material can result in column densities as high as a few 10$^{23}$ cm$^{-2}$, consistent with the values implied by our observations.  Therefore, it is possible that the accretion disk plays some role in absorbing the X-rays from the boundary layer.

The existence of a high column density absorber in the system offers a way to reconcile a high accretion rate with a low observed X-ray luminosity.  Correcting only for interstellar absorption, the luminosity of the hard component in the 0.3--10 keV range is 1.1 $\times$ 10$^{32}$ ($D$/1.6 kpc)$^{2}$ erg s$^{-1}$.  However, when we also correct for the presence of the complex absorber, the luminosity in the same range becomes 1.8 $\times$ 10$^{33}$ ($D$/1.6 kpc)$^{2}$ erg s$^{-1}$--- a  factor of 10 increase.  Our limits on an optically thick component of the boundary layer show that the unabsorbed luminosity of the system could be as high as 1.6 $\times$ 10$^{35}$ ($D$/1.6 kpc)$^{2}$ erg s$^{-1}$.  Therefore, the high degree of intrinsic absorption in the system can easily reduce the emergent X-ray flux by a factor of 1000.  This is one factor that explains the so-called ``missing boundary layer" problem in RS Oph.

\subsection{A low temperature accretion component}
Given the high mass of the white dwarf in RS Oph, the low maximum temperature of the cooling flow suggests that some other cooling mechanism besides thermal bremsstrahlung cooling is at work.  Models of optically thin, radial accretion onto white dwarfs predict that the maximum temperature of the post shock gas is set by the freefall velocity, and therefore the mass, of the white dwarf as follows:
\begin{equation}
T_{\rm{shock}} = \frac{3 \mu m_{\rm{H}} G M_{\rm{WD}}}{8 k R_{\rm{WD}}} 
\end{equation} where $\mu$ is the mean molecular weight (assumed here to be 0.63), $m_{H}$ the mass of one hydrogen atom, and M$_{WD}$ and R$_{WD}$ the mass and radius of the white dwarf respectively.  For accretion through a disk, the velocity of the pre-shock gas is given by the innermost Keplerian orbit, and is a factor of $\sqrt{2}$ less than in the free fall case.  Therefore the resulting temperature, which varies as $v^{2}$, is a factor of two lower.  Assuming $M_{\rm{WD}}$ = 1.3 M$_{\odot}$ and $R_{\rm{WD}}$ = 3 $\times$ 10$^{8}$ cm for RS Oph, the expected $T_{\rm{shock}}$ is $\sim$63 keV.  Cooling flows with high maximum temperatures are seen in several other symbiotic stars believed to harbor white dwarfs as massive as the one in RS Oph (e.g. RT Cru \citep{Luna07} and T CrB \citep{Luna08}).  However, the maximum temperature of the flow in RS Oph derived from our model fits ($<$ 6 keV where the statistics are good) is much lower than in those systems.  We note that the mass accretion rates implied by the cooling flow normalization in our models are of the same order as those in T CrB and RT Cru, i.e a few 10$^{-9}$ M$_{\odot}$ yr$^{-1}$.  Why is the maximum temperature so much lower?

One possible explanation is that cooling by Compton scattering is important in the post shock accretion flow in RS Oph.  In this scenario, there is an intense photon field that Compton scatters off, and cools, the hot post-shock electrons.  The overall plasma temperature will drop if the energy losses due to inverse Compton scattering dominate over those due to free-free emission. \citet{Frank02} (hereafter FKR02) give the following expression for the ratio of the Compton cooling time to the free-free cooling time (their Equation 6.56):
\begin{equation}
\frac{t_{Compt}}{t_{ff}} = \frac{7.5 \times 10^{-5}N_{e}}{U_{rad}T_{e}^{1/2}}
\end{equation}
where $N_{e}$ and $T_{e}$ are the electron density and temperature of the emitting plasma and $U_{rad}$ is the radiation energy density of the scattering light source.  Assuming that the seed photons arise from some source on the surface of the white dwarf, $U_{rad}$ can be approximated as follows.
\begin{equation}
U_{rad} \simeq \frac{L_{rad}}{4 \pi R_{WD}^{2} f_{rad} c} 
\end{equation}
Here, $L_{rad}$ is the luminosity of the radiation source providing the scattering photons,  $f_{rad}$ is the fraction of the white dwarf surface over which this light is emitted, and $c$ is the speed of light.  The post shock density is determined by the mass continuity equation and the strong shock jump conditions so that 
\begin{equation}
N_{e} = \frac{\dot{M}}{4\pi m_{e} R_{WD}^{2} f_{acc} v_{K}} 
\end{equation}
where $m_{e}$ is the electron mass, $f_{acc}$ is the fraction of the white dwarf surface over which mass is accreted (note that this is not necessarily the same as $f_{BL}$, the fraction of the white dwarf surface taken up by an optically thick boundary layer, in Equation 7), and $v_{K}$ is the innermost Keplerian velocity.  Inserting equations 9 and 10 into Eq. 8, expressing $T_{e}$ using Equation 7 in the Keplerian case, and noting that $v_{K}$ = $(GM_{WD}/R_{WD})^{0.5}$, we find that the ratio of the cooling times can be expressed as follows:
\begin{equation}
\frac{t_{Compt}}{t_{ff}} = 2.3~ L_{rad, 33}^{-1} \dot{M}_{-9} R_{9}^{-0.5}  M_{1.3}^{-1.5} f_{rad} f_{acc}^{-1}.
\end{equation}
Here, we have expressed the luminosity of the radiation source, L$_{rad, 33}$, in units of 10$^{33}$ erg s$^{-1}$.  $\dot{M}_{-9}$ is the mass accretion rate onto the white dwarf in units of 10$^{-9}$ M$_{\odot}$ yr$^{-1}$, $R_{9}$ is the radius of the white dwarf in units of 10$^{9}$ cm, and M$_{1.3}$ is the white dwarf mass in units of 1.3 M$_{\odot}$.

In order for Compton cooling to be important in RS Oph, we require that  $t_{Compt}$/$t_{ff}$ $\leq$ 1.  If we assume that the post shock flow is optically thin, and that the low temperature \textit{is} due to Compton cooling, then the cooling flow model will reflect the correct accretion rate, but have a maximum temperature representative of the region in the flow where the density increases to the point where free-free emission once again dominates the cooling.  Therefore, we can take $\dot{M}$  from the cooling flow models developed to describe the data.  The other parameters, $f_{rad}$ and $f_{acc}$, are unknown.  $f_{acc}$ depends on the structure of the inner disk and boundary layer, which is model dependent.  An estimate can be made by calculating the ratio of the scale height of the boundary layer to the radius of the white dwarf.  \citet{Patterson85} give an expression for the boundary layer scale height, which we have expressed in terms of our white dwarf parameters and accretion rates as follows:
\begin{equation}
H = 5.1 \times 10^{7} M_{1.3}^{0.1} \dot{M}_{-9}^{-0.5} ~{\rm cm},
\end{equation} where M$_{1.3}$ and $\dot{M}_{-9}$ have the same meaning as in Equation 11.  We find that on day 537, the average $\dot{M}$ obtained from the cooling flow component in our best fit models is $\sim$2 $\times$ 10$^{-9}$ M$_{\odot}$ yr$^{-1}$, resulting in a corresponding $f_{acc}$ of 0.1 (we concentrate on this epoch since the day 744 \textit{XMM-Newton} observation does not allow us to place constraints on the accretion rate).  The fraction of the white dwarf surface which emits the scattering photon field, $f_{rad}$, depends on the physical origin of the radiation.  

One source could be residual photospheric emission from the white dwarf, in which case $f_{rad}$ $\simeq$ 1.  During a nova outburst, the outer layers of the white dwarf are heated by the nuclear burning taking place on the surface.  Once burning ceases, the white dwarf must radiate away this excess heat.  In work exploring the evolution of a classical nova through a complete accretion-outburst cycle, \citet{Prialnik86} found that shortly after nuclear burning ends, the white dwarf experiences a dramatic drop in luminosity by a factor of 10 as the nuclear ashes cool, and then a longer phase of luminosity decline as the white dwarf loses the excess heat in its outer layers (see their Figure 8).  During that phase, which lasts about 10 years, $L$ $\propto$ $t^{-1.14}$, where $t$ is the time since the onset of the outburst.  For this radiation source to cool the post shock plasma in RS Oph, we find that $L_{rad}$ $>$ 8.4 $\times$ 10$^{34}$ erg s$^{-1}$.  This implies an effective temperature of the white dwarf photosphere of $\sim$190,000 K.  Taking the post-burning luminosity of the white dwarf given in Prialnik (1986) of ~10$^{36}$ erg s$^{-1}$, then at 1.5 years after the outburst we expect the residual luminosity to be of order 10$^{35}$ erg s$^{-1}$.  This would provide the required radiation field to cool the post shock accretion flow.  We note that such a source would be undetectable in our data, even if it is only absorbed by the interstellar medium.

Alternatively, if the boundary layer is partially optically thick, then this additional component could provide the necessary radiation field to cool the optically thin portion of the post shock plasma.  If this is the case, then $f_{rad}$ would be determined by the structure of the optically thick part of the boundary layer.  As we discussed in Section 4.3, an optically thick boundary layer is expected to cover $\sim$10\% of the white dwarf surface.  Again taking an average $\dot{M}$ in the optically thin part of the boundary layer of 2 $\times$ 10$^{-9}$ M$_{\odot}$ yr$^{-1}$, then an $f_{acc}$ of 0.1 requires $L_{rad}$ $>$ 8.4 $\times$ 10$^{33}$ erg s$^{-1}$ for cooling due to Compton scattering to be dominant.  The limits we placed on the existence of an optically thick boundary layer show that such a source of radiation could be present without being detected directly in the observed spectrum of the system.

We acknowledge that invoking Compton scattering as the explanation for the low cooling flow maximum temperature requires a source of radiation which we have not directly observed in the data.  Therefore we also consider a scenario that does not require the presence of such a radiation field.  It may be that RS Oph falls into a range of accretion rates where the transition from an optically thin to an optically thick boundary layer takes place.  \citet{Popham95} explore the properties of optically thick boundary layers for white dwarfs with masses in the range 0.6--1.0 M$_{\odot}$, particularly their dependence on accretion rate, white dwarf mass and rotation rate.  They find that the transition from optically thin to thick occurs at higher accretion rates for more massive white dwarfs (see their Figure 8, upper panel). They note that their studies indicate a transition region between the two states, where the boundary layer is optically thin, but has not yet become hot.  For a one solar mass white dwarf, this would occur for $\dot{M}$ between 3 $\times$ 10$^{-10}$ and 10$^{-9}$ M$_{\odot}$ yr$^{-1}$.  However, the transition should occur at higher $\dot{M}$ for a more massive white dwarf.  Unfortunately, the authors do not elaborate on this regime further - for now we just note that this could offer an explanation for the lack of a very high temperature component in the RS Oph spectrum.  

Any of these scenarios can potentially explain the difference between RS Oph and T CrB, another recurrent nova and symbiotic star.  The longer interval between nova outbursts in T CrB than in RS Oph implies a lower accretion rate in T CrB.  If this is the case, then these two systems may straddle the transition zone discussed in \citet{Popham95}, thus explaining the different cooling flow maximum temperatures. A slightly lower accretion rate may also mean that T CrB has either no optically thick boundary layer component, or if one is present, it may not be luminous enough to result in $t_{Compt}$/$t_{ff}$ $<$ 1.  If the white dwarf photosphere is the main source of Compton scattering photons in RS Oph, then the longer time since the last outburst in T CrB (60 years vs. 2 years) and the lower expected photospheric luminosity may result in less efficient Compton cooling of the post shock flow, and the higher observed maximum temperature.   A potential test of this hypothesis is to look for a low cooling flow maximum temperature within the first few years of the next outburst of T CrB.

\subsection{The mass accretion rate onto the white dwarf}

At each epoch, the spectral models we have developed allow us to explore the range of possible accretion rates onto the white dwarf.  As we discussed in Section 1, existing estimates of the accretion rate in RS Oph using X-ray observations are too low to account for the observed outburst frequency.   Using the results of \citet{Yaron05}, the triggering mass required to power an outburst every $\sim$20 years is 3.5 $\times$ 10$^{-6}$ M$_{\odot}$ for a 1.2 M$_{\odot}$ WD, and 3.47 $\times$ 10$^{-7}$ M$_{\odot}$ for a 1.4 M$_{\odot}$ WD.  The associated accretion rates are therefore 1.8 $\times$ 10$^{-7}$ and 1.7 $\times$ 10$^{-8}$ M${\odot}$ yr$^{-1}$, respectively.  For each dataset, we have found two equally well fitting models.  In the first, the hard component is a single temperature plasma, and in the second it is a cooling flow.  

We can estimate the accretion rate from the single temperature model by obtaining the unabsorbed X-ray flux of the hard component, converting this to a luminosity, and inserting into Equation 1 (again assuming a 1.3 M$_{\odot}$ white dwarf with radius 3 $\times$ 10$^{8}$ cm).  Note that the luminosity depends on the distance to RS Oph, and there is a degree of uncertainty in this quantity.  Throughout this paper, we have assumed a distance of 1.6 kpc (see Section 4.1).  However, larger distance estimates exist in the literature. \citet{Rupen08} proposed $D$ =  2.45 $\pm$ 0.4 kpc, based on the size of the radio outflow spatially resolved with the Very Long Baseline Array (VLBA) during the 2006 outburst.  \citet{Schaefer09} favors an even larger distance, $\sim$4 kpc, based on an average of a number of different methods.  However, at this distance the nebular expansion observed in both the radio and the optical with the Hubble Space telescope \citep{Ribeiro09} imply expansion velocities of $>$8000 kms$^{-1}$, which appears to be too large for a nova. For this reason, we take the larger value of Rupen et al. as the upper limit to the distance.

We obtain the unabsorbed flux of the single temperature hard plasma in XSpec by setting the normalization of the softer plasma components, and all absorption components, to zero.  An associated uncertainty is derived by looking at how the flux changes within the 90\% confidence interval range for both the plasma temperature and normalization.  For the \textit{Chandra} observation, we find a range of unabsorbed 0.3--10 keV fluxes for this component of 0.6--2.0 $\times$ 10$^{-11}$ erg s$^{-1}$ cm$^{-2}$ for both sets of abundances (i.e. models CA and CA-va in Table 4).  Assuming a distance of 1.6 kpc, these flux measurements imply accretion rates of 1.1--3.1 $\times$ 10$^{-10}$ ($D$/1.6 kpc)$^{2}$ M$_{\odot}$ yr$^{-1}$. In the {\it XMM-Newton} observation the flux values are a factor of $\sim$10 lower (0.3--3.2 $\times$ 10$^{-12}$ erg s$^{-1}$ cm$^{-2}$), and hence so is the implied accretion rate (0.6--4.5 $\times$ 10$^{-11}$ ($D$/1.6 kpc)$^{2}$ M$_{\odot}$ yr$^{-1}$).  Placing RS Oph at 2.45 kpc (the Rupen et al. distance) increases the X-ray luminosity, and hence accretion rate, by a factor of 2.3.  Thus, the maximum accretion rate is 7.1 $\times$ 10$^{-10}$ M$_{\odot}$ yr$^{-1}$ 1.5 years after outburst, and 10$^{-10}$ M$_{\odot}$ yr$^{-1}$ 2 years after outburst.

The cooling flow model provides a direct estimate of the mass accretion rate onto the white dwarf, if we assume that the boundary layer is completely optically thin (if an optically thick region exists the accretion rate can be higher, as we'll discuss shortly).  The values found for the \textit{Chandra} observation are in the range 0.7--5.3 $\times$ 10$^{-9}$ M$_{\odot}$ yr$^{-1}$---factors of 3--50 times larger than those inferred from the single temperature plasma model.  Accounting for the uncertainty in the distance, the accretion rate implied by the cooling flow fits can be as large as 1.2 $\times$ 10$^{-8}$ using the radio distance of Rupen et al. This value is within a factor of 2 of the rate required to power an outburst in the system every $\sim$20 years, if the white dwarf is close to 1.4 M$_{\odot}$.  The estimate obtained for the \textit{XMM-Newton} observation is an order of magnitude lower, although with much larger uncertainties given the poorer statistics of that data.  In any case, it seems clear from the variations in X-ray brightness between the two epochs that the accretion rate is variable in quiescence by up to a factor of 10.

If the boundary layer is partially optically thick, which may be the case in RS Oph as we discussed above, then the data are consistent with an even higher mass accretion rate.  As we showed in Section 4.3, a boundary layer with luminosity up to 1.6 $\times$ 10$^{35}$ ($D$/1.6 kpc)$^{2}$ erg s$^{-1}$ is consistent with the \textit{Chandra} data, assuming our best fit model is the correct one.   This implies an accretion rate of 8.5 $\times$ 10$^{-9}$ ($D$/1.6 kpc)$^{2}$ M$_{\odot}$ yr$^{-1}$ through an optically thick component of the boundary layer.  Adding this to the $\dot{M}$ estimate from the cooling flow, and placing the system at 2.45 kpc results in an accretion rate can power an outburst every 20 years.

We can use the UV flickering to place a further constraint on the accretion rate 2 years after the outburst.  Using the values of the flux of the flickering source ($F_{f}$) presented in Table 2, a lower limit to $\dot{M}$ (since we have not performed any bolometric correction and therefore have not accounted for all the flux) can be obtained using the following equation
\begin{equation}
\dot{M} > \frac{8\pi D^{2} R_{WD}}{G M_{WD}} ~F_{f} 
\end{equation}  
Again assuming a distance to RS Oph of 1.6 kpc, a white dwarf mass $M_{WD}$ of 1.3 M$_{\odot}$, and a white dwarf radius $R_{WD}$ of 3 $\times$ 10$^{8}$ cm, we can express $\dot{M}$ in a more convenient form 
\begin{eqnarray}
\dot{M} > 2.3 \times 10^{-10}  \left(\frac{D}{1.6 ~\mbox{kpc}}\right)^{2} \left(\frac{R_{WD}}{3 \times 10^{8} ~\mbox{cm}}\right)  \times \nonumber \\
\left(\frac{M_{WD}}{1.3 ~\mbox{M}_{\odot}}\right)^{-1} \left(\frac{F_{f}}{1.5 \times 10^{-11} ~\mbox{erg~s$^{-1}$~cm$^{-2}$}}\right) ~{\rm M_{\odot} ~yr^{-1}}.  \nonumber \\
\end{eqnarray}
We present the values of $\dot{M}$ at this epoch for our range of assumed reddening in Table 2.  We favor an E(B-V) for the UV light of 0.7 , since this value was derived from earlier UV observations obtained with \textit{IUE} during both outburst and quiescence \citep{Snijders87}.  This E(B-V) value corresponds to an N(H) = 3.7 $\times$ 10$^{21}$ cm$^{-2}$, assuming a conversion factor N(H) = 5.3 $\times$ 10$^{21}$ E(B-V) cm$^{-2}$ \citep{Predehl95}.  This value is higher than the ISM absorption, most likely due to intrinsic attenuation of the UV flux by the red giant wind.  For this value of the reddening, the corresponding lower limit to the mass accretion rate is 2.6 $\times$ 10$^{-9}$ M$_{\odot}$ yr$^{-1}$.  This is much higher than the accretion rate estimates given by Orio (1993) and Mukai (2008), and is much closer the the expected accretion rate given the frequency of observed outburst, as discussed in \citet{Yaron05}.

Ultimately, we have shown that if the hard X-ray emission originates in a cooling flow, then the implied accretion rate is very close to the theoretical rate given by nova theory if the mass of the white dwarf is within $\sim$10\% of Chandrasekhar limit.  If the boundary layer is partially optically thick, then the data are consistent with the high accretion rate if RS Oph is at the distance of Rupen et al. (2.45 kpc).  Since the model fits suggest that the accretion rate is variable in quiescence, then it is possible that at other times the accretion rate is higher.  Thus,  when averaged over the inter-outburst time period, the required $\dot{M}$ is reached.  If the cooling flow model is incorrect, then we must seek some other mechanism to supply the white dwarf with mass, since the rates inferred from the single temperature plasma models are much too low (even at the maximum distance) to result in the observed outburst frequency.  This would most likely be an optically thick boundary layer, which is hidden from our view by the ISM and the intrinsic absorber in the system.

\subsection{Emission from the nova shell}
The soft component is generally described by a two temperature plasma with non-solar abundances.  The temperatures and relative fluxes vary slightly between our models, but the presence of a less absorbed, low temperature plasma is evident from a direct examination of the data (since such soft flux is at odds with the large Fe K edge observed at 7.1 keV).  This component is observed to fade between the \textit{Chandra} and \textit{XMM-Newton} observations, but the temperatures agree within the uncertainties.  Therefore, it is likely that this component arises in the shocked ejecta and red giant wind created during the early stages of the outburst.  The derived emission measures (obtained from the normalization constants of the thermal plasma components, assuming a distance of 1.6 kpc) are in the range 10$^{54}$--10$^{55}$ cm$^{-3}$---a factor of 1000--10000 lower than observed immediately after the outburst \citep{Sokoloski06}.  The cooling time for a shocked optically thin plasma can be calculated by dividing the total thermal energy by the energy lost per unit volume
\begin{equation}
t_{\rm{cool}} = \frac{\frac{3}{2} n_{e} k T_{\rm{max}}}{n_{e}^{2}\Lambda(T)}
\end{equation}
where n is the electron density, T$_{\rm{max}}$ the maximum post shock temperature and $\Lambda\rm{(T)}$ the radiative cooling function at a given T.  Over the range of temperatures where the plasma emits in the X-rays, the cooling function is approximately constant and has a value of $\sim$10$^{-23}$ erg s$^{-1}$ cm$^{3}$ \citep[see e.g. ][]{Raymond77}.  Taking T$_{\rm{max}}$ = 10$^{8}$ K and an estimate of $n_{e}$ in the envelope of the red giant obtained 2--3 yrs after the 1985 outburst of 10$^{7}$ cm$^{-3}$ \citep{Anupama89}, we find a cooling time of the order of 10 yrs (this $n_{e}$ estimate is of the same order of magnitude as many other symbiotic stars as demonstrated in \citet{Luna05}). Thus, it seems reasonable that the shocked material would still be emitting at the times of the new observations.  The two temperature fit is most likely an approximation to the actual temperature distribution of the plasma, which should be continuous.  Our best fit \textit{XMM-Newton} model can account for the \textit{ROSAT} emission observed 7 years after outburst, which suggests that the shell emits for a large fraction of the inter-outburst time.  The soft count rate variation observed between the first and second \textit{ROSAT} observations is larger than that seen over the course of the \textit{Chandra} observation (factor of 3 for \textit{ROSAT} versus 2 for \textit{Chandra}), but they may both be due to the same effect i.e. some combination of changing absorption of the hard component.

The new observations also allow us to examine the long term evolution of the system post outburst.  The X-ray flux has continued to decrease since the end of the supersoft phase, and is a factor of $\sim$2 lower in the energy range 0.3--10 keV in the day 537 observation than in the \textit{XMM-Newton} observation obtained 239 days after the outburst .  By the time of the \textit{XMM-Newton} pointing on day 744, the flux is a factor of $\sim$6 lower than on day 239.  This change of flux occurs primarily in the soft component.  \citet{Ness09} compared the 1.1--1.8 keV flux in all available \textit{Chandra}, \textit{XMM-Newton} and \textit{Swift} observations obtained after the 2006 outburst (this small energy band was chosen to avoid the supersoft emission which appeared several weeks into the outburst),  They found that after June 2006, the flux decayed as $t^{-8/3}$, where t is the time since outburst.  To see if the new data are consistent with this rate of decline, we compared the observed count rates in the same energy range to the rates predicted by the $t^{-8/3}$ decay, using the 1.1--1.8 keV EPIC-pn count rates on days 205 and 239 of the outburst as our initial values. The observed flux is a factor of 5 and 2 higher at 90\% confidence level than the values predicted for days 537 and 744, respectively.  This is likely due to the reestablishment of accretion---in our models, the harder plasma component is the dominant source of flux above 1.4 keV.

\subsection{Comparison of UV results with optical studies}
Recently, \citet{Zamanov10} published a study of optical flickering in RS Oph, using optical photometry obtained in July 2008 and July 2009.  Based on the degree of flickering in each of their program filters (U, B, V or I), they calculate the flux of the flickering component at multiple wavelengths. They then deredden these values assuming E(B-V) = 0.73.  Finally, they determine the best blackbody fit to the derived flickering spectrum at the two observation epochs.  They find that the source of the optical flickering has a temperature of around 9500 K in both observations, but that the flickering source is significantly brighter in 2009 than in 2008, with a bolometric luminosity of $\sim$150 $L_{\odot}$ (vs. 50 $L_{\odot}$ in 2008).  These values correspond to accretion rates (assuming the same white dwarf parameters as we did in Section 4.3) of 3.2 and 1.5 $\times$ 10$^{-8}$ M$_{\odot}$ yr$^{-1}$ assuming a distance of 1.6 kpc.  

Although our UV observations were carried out before both of these observing epochs, it is useful to compare our measured flux with their best fit models of the optical flickering source.  To facilitate this comparison, we note that Zamanov et al. define the flickering flux, $F_{f}$ as $\bar{F}$ - $F_{min}$, whereas we define $F_{c}$ = $F_{max}$ - $F_{min}$, resulting in a difference in flux values of a factor of $\sim$2.  Assuming the same degree of reddening as Zamanov et al. (i.e. E(B-V) = 0.73), we obtain a flickering source flux density from our OM UVM2 (2310 \AA) photometry of 5.3 ($\pm$3.2) $\times$ 10$^{-13}$ erg s$^{-1}$ cm$^{-2}$ \AA$^{-1}$.  This is higher than the value predicted by their best fit to the optical data in July 2008, and consistent with the July 2009 fit (see their Figure 5). The temperature found for the optical flickering source is much too low to be associated with the boundary layer, and is more likely to be a signature of the accretion disk.  

\section{Conclusions}
\label{conclusions}
We have analyzed X-ray and UV data from two observations of RS Oph carried out 537 and 744 days after the 2006 outburst.  The \textit{Chandra} observation reveals that the X-ray emission is variable on short timescales, with the changes most apparent at energies between 1 and 2 keV.  A comparison of the two datasets also reveals long term variability, with the 0.3--10 keV flux reduced by a factor of 3 in the \textit{XMM-Newton} observation.  When compared with the long term brightness evolution observed with Swift, we find that RS Oph is brighter than the trend observed in the Swift data would predict, indicating the existence of a new source of flux in the system, most likely renewed accretion.  The X-ray spectra are well described by two components - a soft optically thin plasma component with temperatures 0.15--0.65 keV, absorbed only by the interstellar medium, and a harder, highly absorbed (N(H) $\sim$ 10$^{24}$ cm$^{-2}$, 97-99\% covering fraction) component, which is equally well described by a single temperature plasma or a cooling flow.  The soft component is observed to decrease in brightness between the two observations, although the temperature remains constant within our measured uncertainties.  We propose that this emission arises in the shocked nova ejecta, which is still cooling from its peak post-outburst temperature, and can remain an X-ray emitter for up to a decade after the outburst.  The hard component arises in the accretion disk boundary layer.  

Our best fit models can also be used to describe the emission observed with \textit{ROSAT} 6 and 7 years after the 1985 outburst, and with \textit{ASCA} 12 years after the outburst.  The current data have brighter soft emission than was observed with either satellite in the quiescence period before the 2006 outburst.  This is most likely because in those observations the nova shell emission, which dominates below 1.4 keV, was much fainter due to the larger amount of time passed since outburst.  The models presented here for both \textit{Chandra} and \textit{XMM-Newton} can be consistent with the hard emission observed with \textit{ASCA}, primarily due to the poor statistics of that dataset.  \textit{ROSAT} was not particularly sensitive to the accretion component, which is most obvious above 2 keV.  However, the differences in count rate observed with \textit{ROSAT} could also be explained by a change in the hard component, as we saw during the \textit{Chandra} observation on day 537.

The temperature associated with the hard component is only 2.6$^{+0.7}_{-0.4}$ keV (for the single temperature plasma) or 5.9$^{+3.5}_{-1.6}$ keV (for the cooling flow model). This is surprisingly low for accretion onto such a massive white dwarf, and strikingly different from the maximum temperatures found in similar systems like T CrB.  We can conceive of two explanations for this low temperature value.  The first is that cooling via Compton scattering dominates the accretion flow, in the boundary layer, lowering the maximum temperature of the emission.  The scattering radiation field could originate in either the white dwarf photosphere, which was recently heated by the nuclear burning associated with the nova outburst, or from some optically thick region in the boundary layer.  The absence of such a source in T CrB (either due to a lower accretion rate or the longer time since the last nova outburst) would explain the difference in observed maximum temperature.  The second possibility is that $\dot{M}$ is such that the system lies in the transition region between a completely optically thick and completely optically thin boundary layer.  In this case T CrB and RS Oph straddle the transition accretion rate, and therefore have different spectral characteristics.  

Our model fits to the X-ray spectra obtained on days 537 and 744, and the presence of UV flickering during the day 744 observation with \textit{XMM-Newton}, can be used to estimate a range of reasonable accretion rates onto the white dwarf.  In the X-rays, the high quality  \textit{Chandra} data provide the most robust estimates of $\dot{M}$.  Uncertainties in the accretion rate come from the fact that two physical models describe the data equally well, and also from the range of distances proposed for the system.  Spectral changes over the course of the \textit{Chandra} observation on day 537, and the lower observed 2--10 keV flux on day 744, strongly suggest that the accretion rate onto the white dwarf is variable during quiescence.

If the single temperature model is correct, then the implied accretion rate is in the range  7 $\times$ 10$^{-11}$ to 5 $\times$ 10$^{-10}$ M$_{\odot}$ yr$^{-1}$ for distances of 1.6 and 2.45 kpc, respectively.  This is clearly at odds with the accretion rate required to power an outburst in the system every 20 years.  If on the other hand the cooling flow model is correct, then the accretion rate is in the range 2 $\times$ 10$^{-9}$ to 1.2 $\times$ 10$^{-8}$ M$_{\odot}$ yr$^{-1}$.  Allowing for the presence of an absorbed optically thick component of the boundary layer that is consistent with the observations, the accretion rate can be as high as 3.6 $\times$ 10$^{-8}$ M$_{\odot}$ yr$^{-1}$.  Thus by accounting for the more complex nature of the X-ray emission, we have shown that the long standing discrepancy between the high accretion rate required to power frequent outbursts in RS Oph, and the low observed X-ray flux, can in fact be resolved.

\acknowledgements
We thank the anonymous referee for their detailed comments and suggestions.  Support for this work was provided by NASA through Chandra award GO9-0027X issued by the Chandra X-ray Observatory Center, which is operated by the SAO for and on behalf of NASA under contract NAS8-03060, and through award NNX09AF82G (to J.L.S.). This research made use of data obtained from the High Energy Astrophysics Science Archive Research Center (HEASARC), provided by NASA's Goddard Space Flight Center.  TN and MO acknowledge the support of the NASA-GSFC \textit{XMM-Newton} program.

\bibliography{rsoph2}

\end{document}